\pdfoutput=1
\documentclass[]{AIAA}
\usepackage{amsmath,amssymb} 
\usepackage{bm}
\usepackage{nomencl}
\usepackage{subfig}
\usepackage{booktabs}
\usepackage{doi}
\usepackage{xcolor}
\usepackage[framemethod=tikz]{mdframed}
\usepackage{setspace}
\usepackage{graphicx}

\usepackage[titletoc,title]{appendix}
\usepackage{float}
\newcommand\BibTeX{{\rmfamily B\kern-.05em \textsc{i\kern-.025em b}\kern-.08em
T\kern-.1667em\lower.7ex\hbox{E}\kern-.125emX}}
\usepackage{fancyhdr}
\begin{document}



\title{Quad-rotor Flight Simulation in Realistic Atmospheric Conditions}

\author{Behdad Davoudi\footnote{PhD Candidate.}} 

\author{Ehsan Taheri\footnote{Post Doctoral Fellow, currently Research Assistant Professor, Texas A\&M University.}}

\author{Karthik Duraisamy\footnote{Associate Professor.}}

\author{Balaji Jayaraman\footnote{Visiting Researcher, currently Assistant Professor of Mech. \& Aero. Eng., Oklahoma State University.}}

\author{Ilya Kolmanovsky\footnote{Professor.}} 
\affiliation{Department of Aerospace Engineering, University of Michigan, Ann Arbor, MI 48109}

\newcommand{\degree}{\ensuremath{^\circ\,}} 
\setlength{\nomitemsep}{-\parsep}

\begin{abstract}
In trajectory planning and control design for unmanned air vehicles, highly simplified models are typically used to represent the vehicle dynamics and the operating environment. The goal of this work is to perform real-time, but realistic flight simulations and trajectory planning for quad-copters in low altitude (<500m) atmospheric conditions. The  aerodynamic model for rotor performance is adapted from  blade element  momentum theory and validated against experimental data. Large-eddy simulations of the atmospheric boundary layer are used to accurately represent the operating environment of unmanned air vehicles. A reduced-order version of the atmospheric boundary layer data as well as the popular Dryden model are used to assess the impact of accuracy of the wind field model on the predicted vehicle performance and trajectory. The wind model, aerodynamics and control modules are integrated into a six-degree-of-freedom flight simulation environment with a fully nonlinear flight controller. Simulations are performed for two representative flight paths, namely, straight and circular paths. Results for different wind models are compared and the impact of simplifying assumptions in representing rotor aerodynamics is discussed. 
The simulation framework and codes are open-sourced for use by the community. 

\end{abstract}
\maketitle
\thispagestyle{fancy}%
\section*{Nomenclature}
\noindent
\begin{tabular}{@{}lcl@{}}
%
%

  \bf Symbols && {}\\
    $c$           &=& {Rotor chord, m} \\
    $\bar{c}_\text{d}$       &=& {Lumped drag coefficient } \\
    $C_{d_\text{0}}$       &=& {Profile drag coefficient } \\
    $C_\text{l}$        &=&  {2-D lift coefficient } \\
    $C_{l_\alpha}$ &=& {2-D lift curve slope } \\
    $C_\text{P}$        &=&  {Power coefficient, $P/(\rho \pi R^5 \omega^2)$ } \\
    $C_\text{Q}$        &=&  {Torque coefficient, $Q/(\rho \pi R^5 \omega^2)$ } \\
    $C_\text{T}$        &=&  {Thrust coefficient, $T/(\rho \pi R^4 \omega^2)$}\\
    $\bold{D}$          &=&  {Drag}, Newton \\
    $N_b$         &=&  {Number of blades} \\
    $\textbf{r}$       &=& $[x,y,z]^T$ {inertial position vector, m}\\
    $r$          &=&  {Radial distance of a rotor span-wise station, m} \\
    $R$          &=&  {Rotor radius, m} \\
    $Re$         &=&  {Reynolds number, $Vc/\nu$ } \\
    $T$         &=&  {Total thrust, Newton } \\
    $T_i$         &=&  {Thrust of rotor \# i, Newton } \\
    $\bold{V}$          &=&  {Velocity in inertial frame}, m/s \\
    $\bold{V_{\text{rel}_\text{B}}}$          &=&  {Relative velocity represented in body frame, $\bold{V-V_\text{W}}$} \\
    $\bold{V_\text{W}}$          &=&  {Wind velocity}, m/s \\
    $V_{\text{tip}}$    &=&  {Rotor tip velocity}, m/s \\
    $W$          &=&  {Weight of the quad-copter} \\
    $\alpha$     &=&  {Sectional angle of attack, deg} \\
    $\lambda$          &=&  {Inflow ratio} \\
    $\lambda_c$          &=&  {Climb inflow ratio} \\
    $\lambda_h$          &=&  {Hover inflow ratio} \\
    $\mu$          &=&  {Advance ratio, $\sqrt{V_{\text{rel}_{\text{B}_x}}^2+V_{\text{rel}_{\text{B}_y}}^2}/V_\text{tip}$} \\

    \end{tabular}
    
\begin{tabular}{@{}lcl@{}} 
 $p,q,r$       &=& {Body angular rates, rad/s}\\
    $\Phi$          &=&  {Inflow angle, rad} \\
    $\Psi$       &=&  {Azimuth angle, deg } \\
$\psi, \theta, \phi$          &=&  {Yaw, Pitch and Roll, Euler angles}, rad \\
    $\tau_x, \tau_y, \tau_z$       &=&  {Torque components in body frame - Roll, pith, and yaw torques, respectively} \\
    $\nu$        &=&  {Kinematic viscosity, m$^2$s$^{-1}$ } \\
$\sigma$     &=&  {Blade solidity, (blade area)/(rotor area)}\\
    $\Theta$          &=&  {Blade Pitch, rad} \\
    $\rho$       &=&  {Air density, kg~m$^{-3}$ } \\
    {$\omega$}   &=&  {Rotor angular velocity, rad~s$^{-1}$} \\
\end{tabular} 
\section{Introduction}

Over the past decade, small Unmanned Aerial Systems (sUAS) (or drones) \cite{cai2014survey}  have been increasingly utilized in a variety of tasks including rapid package delivery \cite{lee2017optimization}, border patrolling \cite{wall2011surveillance}, damage inspection \cite{irizarry2012usability}, disaster management \cite{adams2011survey} , mapping \cite{nex2014uav}, precision agriculture \cite{zhang2012application}, etc. With the unprecedented growth in the number of flight vehicles, simulation tools are required for trajectory prediction and validation, especially in the context of certification by analysis.  

\begin{figure}[H]
	\centering
    \includegraphics[width=4in]{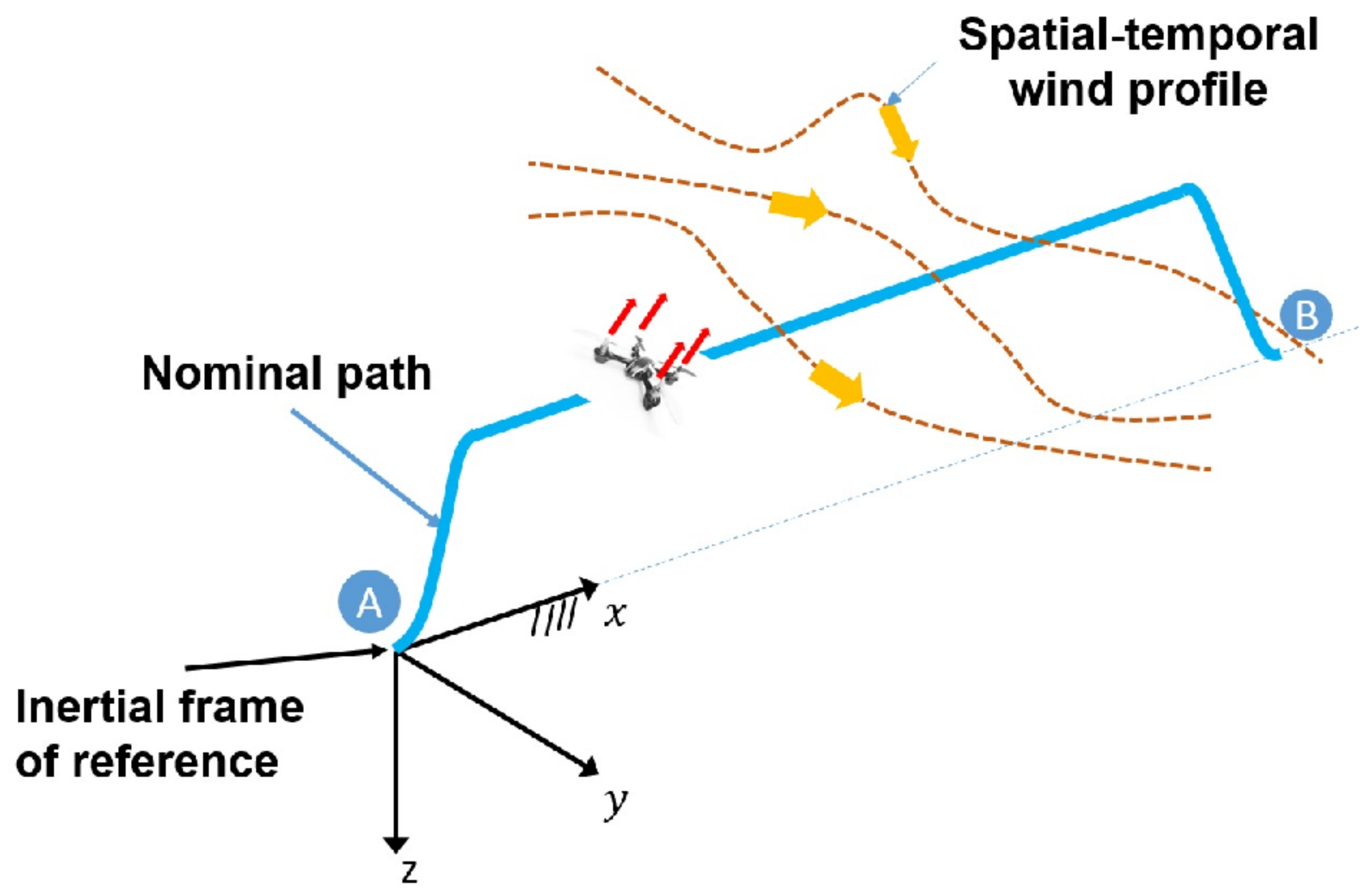}
    \caption{A representative trajectory for a quad-copter and the operating environment.}
	\label{fig:TrajWindSchematic}
\end{figure}

A representative trajectory for a quad-copter is shown in Figure \ref{fig:TrajWindSchematic} where the vehicle is assigned, for instance, with a package delivery task. It dispatches from point `A' (say a service center) to point `B' (that may represent a delivery designated area) on a nominal trajectory. 

The quad-copter is subject to extraneous factors such as wind/gusts that affects its performance and trajectory and may deviate its path from the planned one. In practice, there are various along-the-path constraints that have to be taken into consideration during path-planning, specifically, for very low-altitude flight since the quad-copter has to adopt certain maneuvers to avoid colliding with obstacles. 


A comprehensive approach to validating and planning trajectories of flying vehicles requires, in general, a framework that takes into account various aspects including 1) appropriate-fidelity aeromechanical and dynamic models of the vehicle to present a realistic view of the actual motion, 2) accurate modeling of the environment, which may include modeling of external factors that alter the trajectory of the vehicle including wind and gust models, 
3) development of a flight controller to control vehicle trajectory to within a prescribed accuracy, 4) a guidance (motion-planning) algorithm that provides a nominal ``optimal'' trajectory, and 5) a navigation model to represent the sensory data and the measurement noise, which will be used to update the location/orientation of the flying vehicle to be used for guidance and by the flight controller. 

Depending on the stage and/or nature of the study, various simplifications are invoked to facilitate the task of trajectory prediction and performance analysis \cite{berning2018rapid}. For instance, it is common to consider a three-degree-of-freedom model to only trace the trajectory of the center of mass of the vehicle while the orientation of the vehicle is considered to be of secondary importance for early analysis \cite{zipfel2007modeling}. As another example, in launch vehicle or entry, descent and landing trajectory design/analysis, it is common to consider four degrees-of-freedom to trace the motion of the center of mass and the orientation of the body of the vehicle with respect to a reference direction, namely, the pitch angle \cite{mall2016trigonomerization}. A similar approach is used for motion analysis of flying vehicles when only planar motion is assumed. The inclusion of additional degrees of freedom is usually considered for higher-fidelity models and at later stages of design of systems to gain better understanding of the motion and to obtain realistic flight performance envelopes \cite{desai2008six,roshanian2014multidisciplinary,kamyar2014aircraft}. 

Simplified atmospheric models are typically used for mission planning and certification of unmanned air vehicles (UAVs). In practice, the operating environment of UAVs flying at low altitudes ($<500$ m) is not only subject to strong mean velocity gradients (shear), but also involves intermittent unsteady wind gusts that contain a non-trivial fraction of energy compared to the mean flow. Further, the characteristic size of the turbulent eddies, even in a stable boundary layer, is of the order of a few meters \cite{kosovic2000large}, which is similar to the size of an sUAS. Accounting for such scales becomes critical to the vehicle aeromechanics and thus to more accurate trajectory prediction and  validation tasks.

A widely used tool for numerical weather prediction is the Weather Research and Forecasting (WRF) model ~\cite{wrf1,wrf2}. WRF is, however, a  \textit{mesoscale model} and is typically used over large spatial domains ($O(1000)$ km) and with coarse resolutions ($O(2)$ km). WRF thus cannot be relied upon to represent all eddies and gusts of interest for sUAS trajectory prediction and validation purposes, and thus a different, higher resolution approach is required. 

In addition, a critical element in performance and/or trajectory analysis of flying vehicles is to use accurate dynamical models for the propulsion system. For an airplane, this entails modeling of the engine/propellers, whereas for a quad- or multi-copter not only the individual performance of the propellers, but their mutual interactions become important, and have to be taken into consideration for certain types of maneuvers \cite{ventura2018high}. In particular, the key requirement for the aeromechanics model is to provide an effective characterization of the response of a sUAS to unsteady gust loads \cite{sydney2013dynamic}. An additional requirement is that the fidelity of the model should be such that it is easily integrable into the trajectory validation and planning modules and the necessary computations
can be performed in near-real time. 

It is common practice to approximate the actual performance of a propeller (i.e., the thrust or torque) using idealized, simple algebraic models to facilitate numerical analyses. The so-called static models (namely, the thrust and torque of a rotor are expressed in terms of the square of rotor speed) are used extensively for hovering, and slow maneuvers. However, these models do not capture the realistic performance of a propeller, which in general, depends on the inflow velocity. In particular, more accurate models have to be used for flight trajectories that can be utilized in demanding situations such as high speed forward and descent flight phases \cite{huang2009aerodynamics}. In low-altitude flight, these aerodynamic phenomena can influence the vehicle dynamics in the presence of wind and gusts. Towards this end, a propeller thrust model in forward flight was presented in \cite{zhang2014survey} for a blade with linear twist.

 On the other hand, the use of higher-fidelity models such as those based on computational fluid dynamics~\cite{ventura2018high,lakshminarayan2006computational} or even vortex-based methods~\cite{kim2009interactional} to predict the aeromechanics of flight vehicles is computationally demanding, and thus is not feasible in trajectory optimization or trajectory prediction settings. As a consequence, an additional challenge is to establish a set of computationally efficient and effective models of the aeromechanics and flight dynamics of the sUAS vehicles.

The main contribution of this paper is to present efficient and accurate models of atmospheric conditions and vehicle aeromechanics, with the goal of utilizing them in real-time trajectory planning, validation and control. Specifically, the operating environment is characterized via atmospheric boundary layer simulations. Aeromechanical models of appropriate fidelity are derived using momentum and blade element theories with an emphasis on low-altitude flight. These models are integrated with a numerical simulation environment that propagates a six-degree-of-freedom (6DoF) model of a quad-copter along two representative nominal flight trajectory profiles: 1) an ascent-straight-descent profile and 2) a circular path. In addition, the trajectory of the quad-copter is controlled using a fully non-linear backstepping control algorithm. Results from different wind models and propeller models are compared and discussed, highlighting the importance of modeling fidelity for realistic trajectory planning, validation and control.   

The remainder of the paper is organized as follows. Section \ref{sec:wind} introduces the atmospheric boundary layer model. Section \ref{sec:aeromodel} presents a review of the  aerodynamic model. Section \ref{sec:control} discusses the details of a fully nonlinear flight controller (namely, translational-rotational control algorithm). In Section \ref{sec:coupling}, the coupling between the aerodynamic models and control module is presented. Aerodynamic, dynamic, control and wind models are integrated, and flight simulation results for a variety of problems are performed and presented in Section \ref{sec:results}. Section \ref{sec:sum} provides a summary and conclusions.

 

\section{Modeling Atmospheric Gust Effects} \label{sec:wind}
The popular approach to represent atmospheric gusts in aviation applications such as trajectory estimation relies on stochastic~\cite{von1948progress,MIL-STD1990,MIL-HDBK2012} formulations and models of the spectral energy function (and, in turn, the trace of the spectral energy tensor)~\cite{pope2001turbulent}. While computationally efficient, such methods have two major limitations: (i) use of parameterized equilibrium phenomenology that is often inaccurate, and (ii) not explicitly accounting for the structure of the spectral energy tensor. The components of the spectral energy tensor contain information about the anisotropy and coherence in the turbulence structure. In reality, the atmospheric boundary layer (ABL) turbulence is characterized by strong and highly coherent eddying structure that contributes to the uncertainty associated with the predicted trajectory. In this study, we adopt high fidelity Large Eddy Simulations (LES) of the ABL that accurately represent energy containing turbulence eddying structures at scales that are dynamically important for unmanned aerial flight.

The canonical ABL used to generate the wind model data for this study is modeled as a rough flat wall boundary layer with surface heating from solar radiation, forced by a geostrophic wind in the horizontal plane and solved in the rotational frame of reference fixed to the earth's surface. The lower troposphere that limits the ABL to the top is represented with a capping inversion and the mesoscale effects through a forcing geostrophic wind vector. In fact, the planetary boundary layer is different from engineering turbulent boundary layers in three major ways: 1) Coriolis Effect: The rotation of the earth causes the surface to move relative to the fluid in the ABL, which results in angular displacement of the mean wind vector that changes with height, 2) Buoyancy-driven turbulence:  The diurnal heating of the surface generates buoyancy-driven temperature fluctuations that interact with the near-surface turbulent streaks to produce turbulent motions, and 3) Capping Inversion:  A layer with strong thermal gradients that caps the microscale turbulence from interacting with the mesoscale weather eddies. 

\begin{figure}[H]
\centering
\includegraphics[scale=.61]{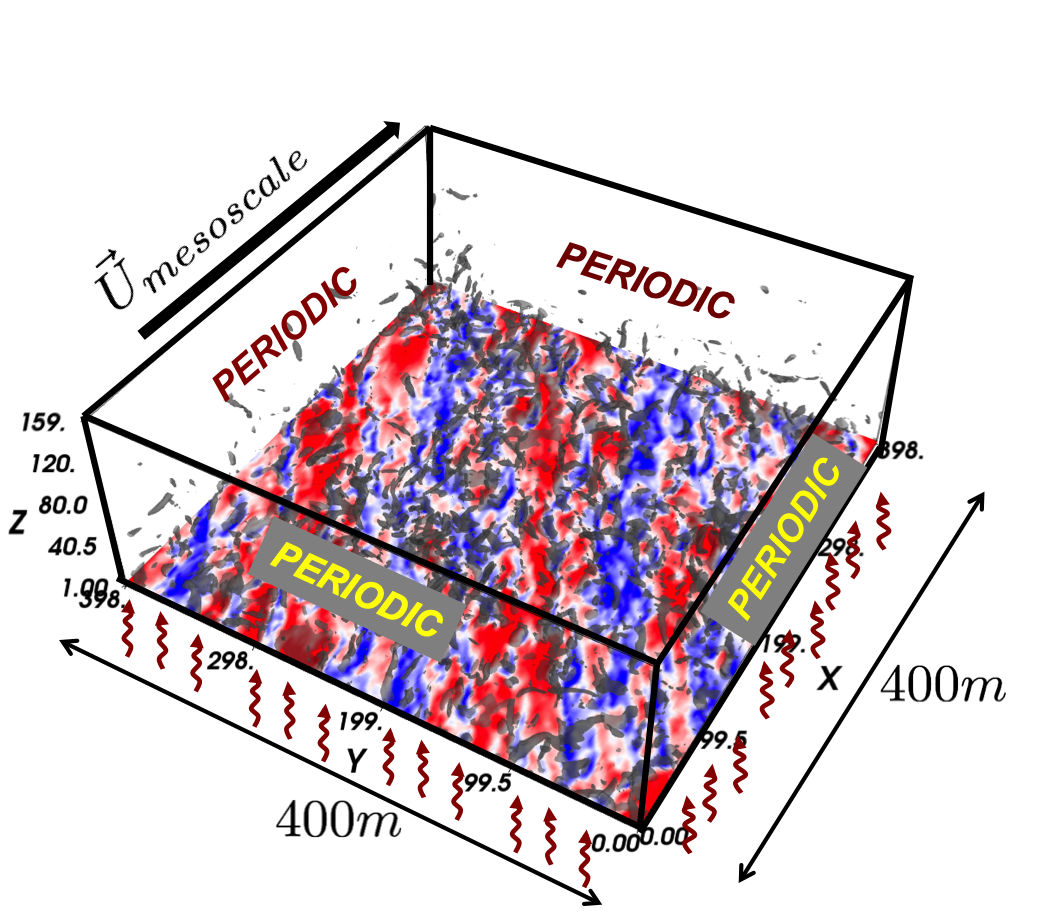}
\caption{Schematic showing the Coriolis effect in a 3D visualization of ABL turbulence for a neutral ABL with $–z_i/L=0$ using LES.}
\label{iso}
\end{figure}
 
In Figure \ref{iso}, the mesoscale wind drives the ABL along the x-direction while the rotation of the earth's surface orients the surface layer turbulence to nearly $\sim 30$ degrees relative to the imposed wind vector (along the streaks). The isosurfaces (grey) show vorticity magnitude at a value of $0.45\ s^{-1}$ and the isocontours show the horizontal fluctuating velocity. The blue regions denote low speed streaks while the red regions represent high speed streaks.

\subsubsection{LES Methodology and Simulation Design:}

The Reynolds number of the daytime atmospheric boundary layer is extremely large. Hence, only the most energetic atmospheric turbulence motions are resolved. The  eddies in the surface layer are highly inhomogeneous in the vertical (z), but are clearly homogeneous in the horizontal direction. LES attempts to resolve to the order of the grid scale, the energy containing eddy structures. 

Using a grid filter, one can split the fluctuating instantaneous velocity and potential temperature into a resolved and sub-filter scale (SFS) components. The canonical, quasi-stationary equilibrium ABL is driven from above by the horizontal mesoscale `geostrophic wind' velocity vector, $\vec{U}_\text{g}$, and the Coriolis force is converted into a mean horizontal pressure gradient oriented perpendicular to $\vec{U}_\text{g}$. In the LES of ABL, the molecular viscous forces are neglected and the surface roughness elements of scale $z_\text{0}$ are not resolved by the first grid cell $(z_\text{0} \ll \Delta z)$. Buoyancy forces are accurately predicted using the Boussinesq approximation. The momentum equation for resolved velocity contains a sub-filter scale (SFS) stress tensor that is modeled using an eddy viscosity formulation with the velocity scale being generated through a 1-equation formulation for the SFS turbulent kinetic energy~\cite{moeng1984large,moeng1986large}. 
The LES equations are shown below in Eqns.~\eqref{eq1}-\eqref{eq3}.
A detailed discussion of the numerical methods is available in ~\cite{khanna1997analysis,jayaraman2014transition,jayaraman2018transition}. In the equations below, $\tilde{u}$ represents the filtered velocity, $\tau^{SFS}$ the subfilter scale stresses, $p^*$ the modified pressure and $\tilde{\theta}$ the filtered potential temperature. 
\begin{align}
\nabla\cdot\tilde{u}&=0, \label{eq1} \\
\frac{\partial\tilde{u}}{\partial t}+\nabla \cdot \left(\tilde{u}\tilde{u}\right)&=-\nabla p^*-\nabla \cdot \tau_u^{SFS}+\frac{g}{\theta_\text{0}}\left( \tilde{\theta}-\theta_\text{0} \right)+f \times \left( u_\text{g} - \tilde{u}\right), \label{eq2} \\
\frac{\partial \tilde{\theta}}{\partial t}+\nabla \cdot \left( \tilde{\theta}\tilde{u}\right)&=-\nabla \cdot \tau_{\theta}^{SFS}. \label{eq3}
\end{align}
While the effects of buoyancy are highly pronounced in ABL turbulence, and significantly impact its structure~\cite{jayaraman2014transition,jayaraman2018transition}, we chose a more benign neutral stratification for this study. To pack sufficient resolution to capture scales relevant to small fixed wind unmanned vehicles, we restricted the domain size to $400m\ \times\ 400m\ \times \ 600m$. The cartesian LES grid has a resolution of $200\ \times\ 200\ \times \ 300$ for a uniform spacing of 2m in each spatial direction. To realistically mimic the interface between the mesoscale and microscale atmospheric turbulence, a capping inversion was specified at a height of 280m. The surface heat flux is set to zero for this neutral ABL simulation and a Coriolis parameter of $f=0.0001s^{-1}$ is chosen to represent continental United States. The bottom surface is modeled as uniformly rough with a characteristic roughness scale of 16cm that is typical of grasslands. The dynamical system described in Eqs.~\eqref{eq1}-\eqref{eq3} is forced by an impossed mean pressure gradient, $\nabla \bar{P}$ usually specified in terms of a geostrophic wind as $\nabla \bar{P}=-f \times {u}_\text{g}$. For this model, $u_\text{g}$ is set to 8m/s which corresponds to a moderately windy day. The equation system is solved using the pseudo-spectral method in the horizontal with periodic boundary conditions and second-order finite difference in the vertical. The time marching is accomplished using a third-order Runge-Kutta method. The computational set-up is exactly as shown in Figure.~\ref{iso}. 
Further details about the computational methods and models can be obtained from ~\cite{jayaraman2014transition,jayaraman2018transition,khanna1997analysis}.



\begin{figure}[H]
\centering
\includegraphics[scale=.5]{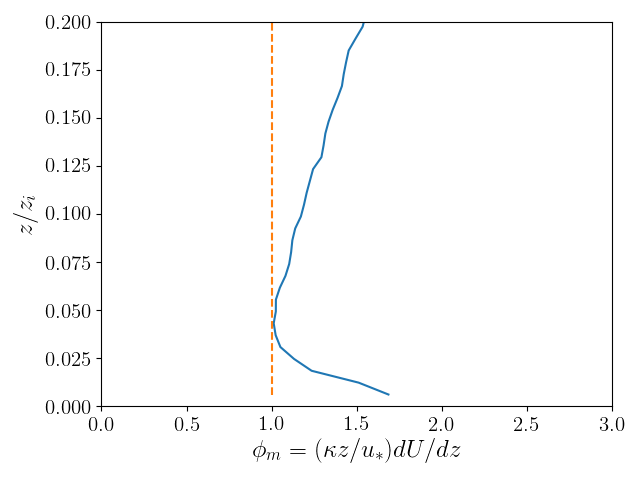}
\caption{Comparison of the non-dimensional mean velocity gradients in the surface layer for neutral ABL with $\kappa=0.4$. For the neutral ABL, the $\Phi_\text{M}$ value is unity. The predicted curve is blue and the expected behavior from phenomenology is orange indicating the existence of the log-law overshoot. 
}
\label{fig3}
\end{figure}

\subsubsection{Validation of Results}
The LES framework has been validated for equilibrium conditions with experimental data~\cite{johansson2001critical} and well known phenomenology~\cite{khanna1997analysis} in the past and is well established as a high fidelity bench tool for modeling near surface atmospheric flows. In this study, we adopt a similar strategy and assess the non-dimensional near-wall scaling from simulation data with respect to the law of the wall, and Monin-Obukhov (M-O) similarity theory arguments for neutral stratification. 
Particularly, we compute the non-dimensional mean gradient $\Phi_\text{M}=\frac{\kappa Z}{u_{\tau}}{d\langle \tilde{u}\rangle}/{dz}$ which should be closer to unity in the inertial logarithmic region of the ABL. In the above, the non-dimensionalization is performed using the appropriate choice of near-surface parameters for length (distance from the wall, $z$) and velocity (friction velocity, $u_{\tau}$). 
For a constant value of the mesoscale wind and surface heat flux, the turbulent flow field evolves into a fully developed equilibrium boundary layer. After verifying the existence of statistical stationarity, converged statistics were plotted. Fig.~\ref{fig3} shows the near-wall variation of $\Phi_\text{M}$ as a function of normalized distance from the surface, ($z/z_\text{i}$), where $z_\text{i}$ is the height of the ABL. We observe that $\Phi_\text{M}$ assumes values closer to unity with deviations known to arise from a combination of inaccuracies in the assumed value of the von Karman constant, $\kappa$, numerical errors and near-wall modeling, which introduces an artificial numerical viscous length scale~\cite{brasseur2010designing}. The LES is considered acceptable as long as these deviations are small as observed in this case.



\subsubsection{Non-stationary ABL Turbulence}

\begin{figure}
	\centering
	\subfloat[Mesoscale winds \label{fig4a}]{%
		\includegraphics[width=0.41\textwidth]{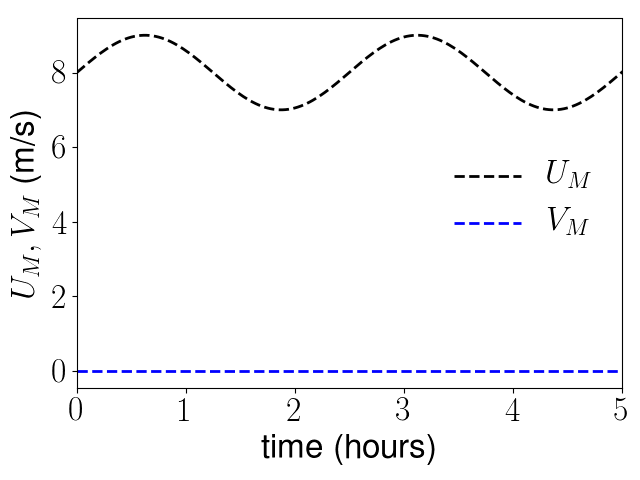}	
	}
	\hfill
	\subfloat[Geostrophic wind\label{fig4b}]{%
		\includegraphics[width=0.42\textwidth]{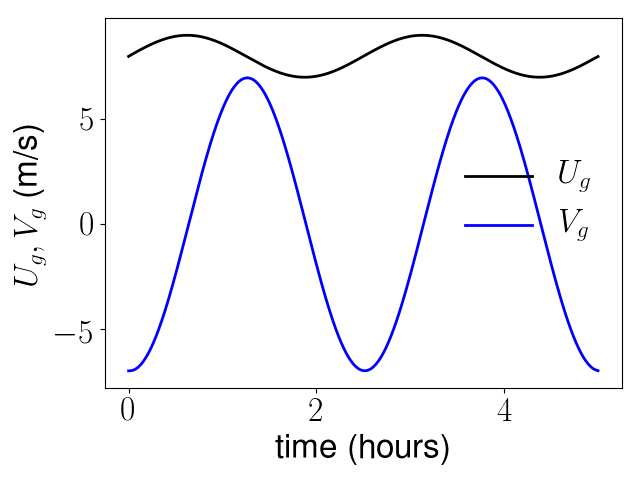}	
	}
	\hfill 
	\subfloat[Friction velocity\label{fig4c}]{%
		\includegraphics[width=0.42\textwidth]{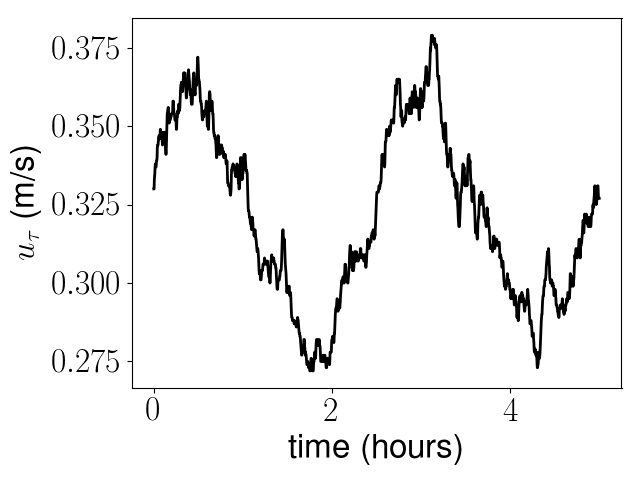}	
	}\hfill
	\subfloat[Surface layer variances\label{fig4d}]{%
		\includegraphics[width=0.42\textwidth]{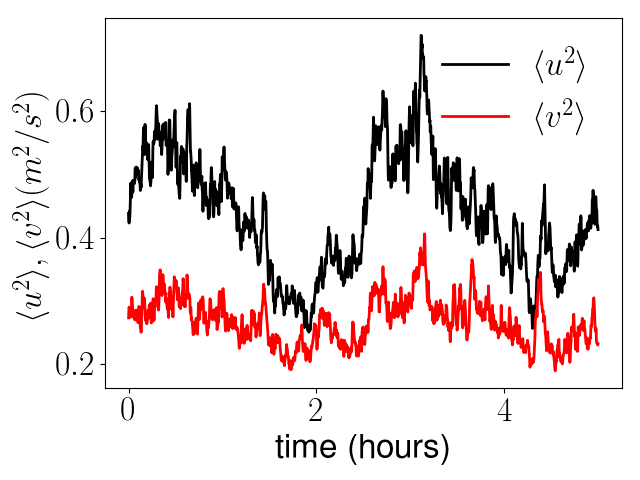}	
	}
	\caption{\label{fig:NonEqABLStats} Non-stationary ABL turbulence and its impact on statistical structure: (a) mesoscale wind vector; (b) geostrophic wind vector; (c) friction velocity and (d) near surface variance of the velocity field. Evidently, the periodic mesoscale wind forcing generates a sinusoidal geostrophic wind vector. The history-dependent response of the turbulent boundary is shown in the form of friction velocity and surface layer variances.   }
\end{figure}

While the above discussion focuses on modeling equilibrium turbulence, the realistic wind fields observed in the atmosphere is often not in equilibrium. Such deviations from equilibrium arise due to multiple reasons including diurnal variations, spatial heterogenieties, terrain and influence of nonstationary mesoscale wind events. To assess the influence of some of these effects on the turbulence structure, we incorporate non-stationary, slow-varying sinusoidal wind (period of O(hours)) on a neutral ABL. This forcing strategy follows the earlier work on mesi-micro coupling by Jayaraman and co-authors in \cite{jayaraman2014nonequilibrium}. The LES design for this simulation is the same as that of the equilibrium ABL LES with the exception of a time-varying mesoscale wind vector. 
Figs.~\ref{fig4a} and \ref{fig4b} show the imposed variations of the mesoscale and geostrophic winds and the resulting ABL turbulence responses are illustrated in Figs.~\ref{fig4c} and \ref{fig4d}. In particular, the friction velocity and the horizontal velocity variances (sampled at $50$m from the surface) are shown to evolve in a non-periodic fashion while the forcing winds are periodic indicative of time history effects and deviations from equilibrium. Such temporally varying statistical structure is not yet understood well for complex flow systems, but can nevertheless impact flight performance of small vehicles, especially when large wind events sweep through an urban canopy resulting in noticeable local gusts.  

\subsubsection{Some sample results}

\begin{figure}[H]
\centering
\includegraphics[scale=0.30]{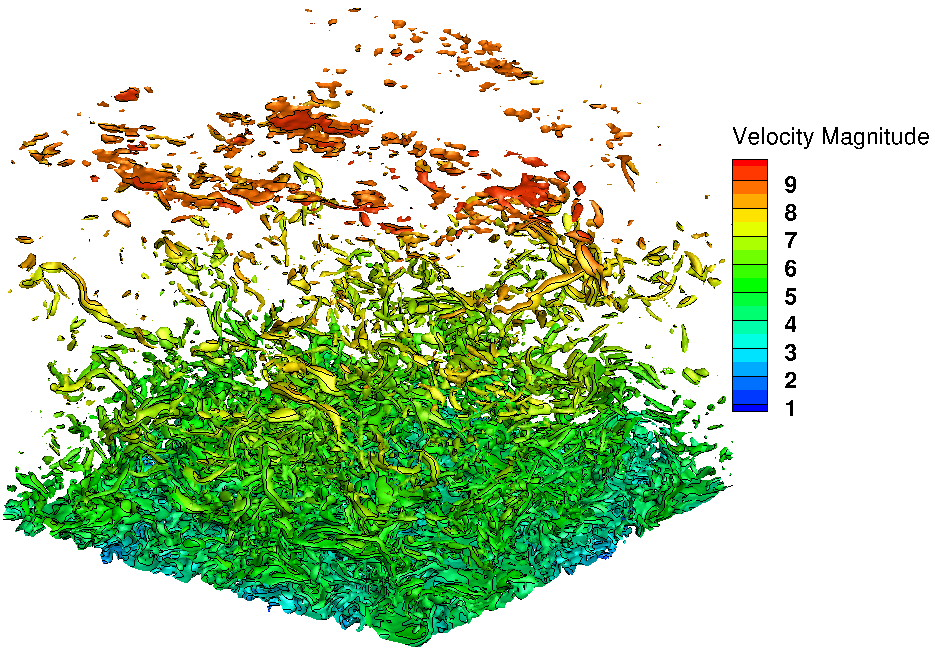}
\caption{Vorticity isosurfaces colored by streamwise velocity.  
}
\label{fig:vor}
\end{figure}

\begin{figure}[H]
\centering
\includegraphics[scale=0.2]{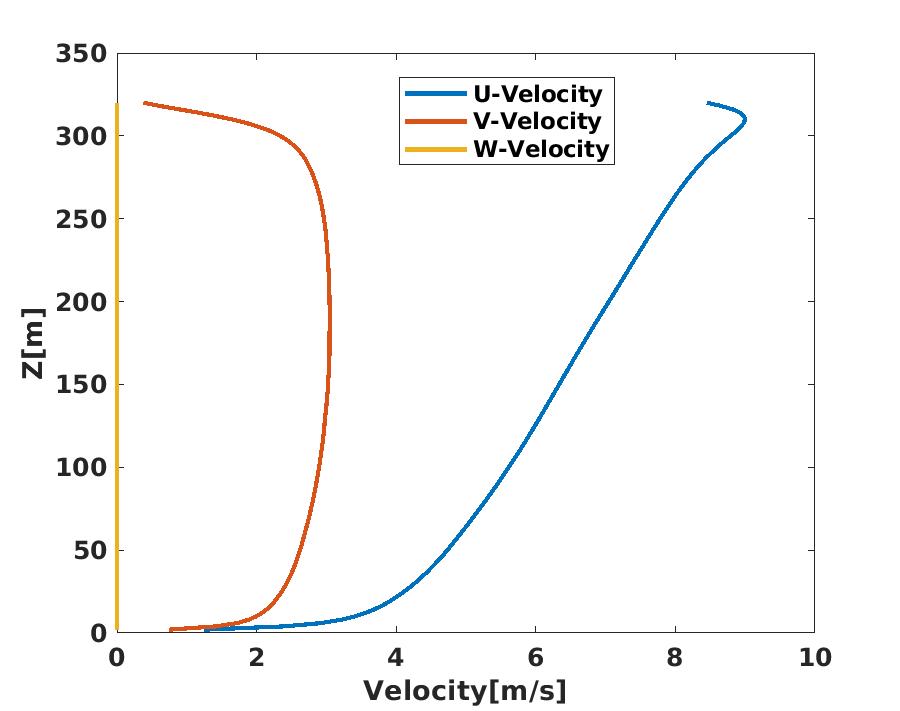}
\caption{Temporally and spatially averaged wind velocities. 
}
\label{fig:UVW_avg}
\end{figure}

 Simulation data were obtained for 100 time instants that were 12.1 seconds apart. For a random snapshot, the vorticity iso-surface colored by U-Velocity is plotted in Fig. \ref{fig:vor}. It is noted there are more disturbances closer to the ground than higher altitudes. The spatially and temporally averaged wind velocity is shown in Fig. \ref{fig:UVW_avg}.

The U-Velocity indicates a boundary layer profile, and there is also a considerable spanwise mean velocity component (y-direction). It is evident that the spatio-temoral average of the W-Velocity is very negligible. However, the W-Velocity can be as high as 1 m/s in magnitude for some time-instances and locations in the field.

\subsection{Reduced-order Wind Representations}
To assess the importance of the details of the wind field, and to reduce the memory requirements, proper orthogonal decomposition~\cite{berkooz1993proper} is  utilized. Given a matrix A of size  $m \times n$, the singular value decomposition is given by:
\begin{equation*}
   A_{\text{m}\times \text{n}}= \hat{U}_{\text{m} \times \text{n}} \Sigma_{\text{m}\times \text{m}} V^{*}_{\text{n}\times \text{n}} 
\end{equation*}
where $\hat{U}, \Sigma$ and $V$ are matrcies of the left singular basis vectors, singular values  and unitary right singular vectors,  respectively. For every component of velocity, the data is stacked into a rectangular matrix. That is, each column  represents  time instances, and each row has the velocity in the three dimensions stacked as a  column vector of length  $Nx \times Ny \times Nz$.



\begin{figure}[H]
\centering
\includegraphics[scale=0.25]{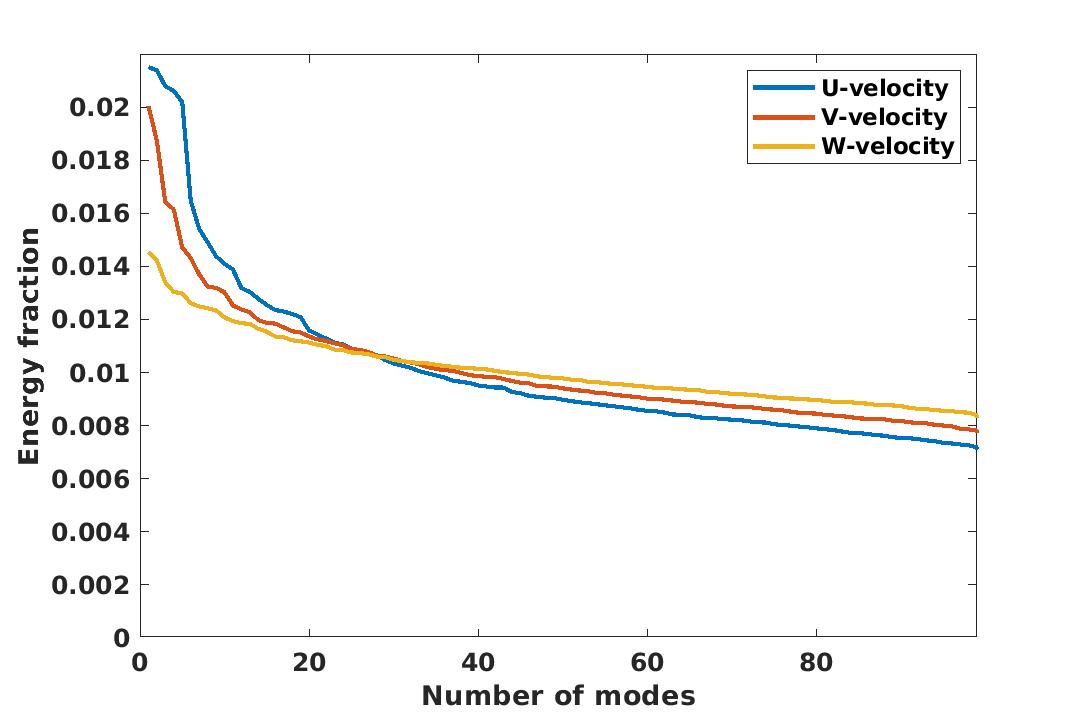}
\caption{Energy contained in each singular mode}
\label{svd}
\end{figure}

The fraction of energy corresponding to each singular value namely: $\frac{\sigma (i) }{\sum \sigma (i)}$ is shown in Figure~\ref{svd}. 
A reduced-order representation can be constructed by utilizing projections on a truncated number of modes $N < 100$. For the wind data, we have chosen $N$ to be 10 modes, and have reconstructed the new wind fields as will be  presented in the Results section.  

\subsection{A benchmark wind model: Dryden turbulence model}

The Dryden wind turbulence model~\cite{hoblit1988gust} uses an empirical spectral representation to add velocity fluctuations to the mean velocity. In  this work, a continuous representation of the Dryden velocity spectra with positive vertical and negative lateral angular rates spectra is used. This representation is based on 
Military Handbook MIL-HDBK-1797B \cite{MIL-HDBK-1797B}. The inputs to the Dryden model are altitude, vehicle velocity (in the inertial reference frame) and direction cosine matrix, and the output is the gust velocity in the body frame. The mean wind velocity is then  added to the fluctuations to represent the full wind. 

The current model provides the mean wind velocity using the  U.S. Naval Research Laboratory $\text{HWM}^{\text{TM}}$ routine. The typical inputs to this model are altitude, longitude, geopotential altitude, and the specific time of interest. The model predictions vary for different spots in the world and time of the year and are of a very low fidelity compared to the ABL simulation. Thus, we have used the ABL data (Section \ref{sec:wind}) to input the wind magnitude and direction for  comparison purposes.

Data from Fig. \ref{fig:UVW_avg} is used to determine the  magnitude of the wind velocity and the wind direction that are input to the aforementioned built-in MATLAB functions. Specifically, the mean wind speed and direction at 6 m are inputs to Dryden Wind Turbulence Model. Those values are approximately: $3.40$ m/s and $240\degree$, respectively. It is noted that the wind direction is measured from the North in a clock-wise positive setting. 





\section{Aerodynamic Model} \label{sec:aeromodel}

In this work, each rotor of a quad-copter is modeled individually using fundamental potential flow theory, while taking into account tools from helicopter rotor aerodynamic modeling. As a prototypical example, Mishra et al.~\cite{mishra2018multiple} utilized an adaptation of extended blade element momentum theory from~\cite{leishmanbook} and validated their  steady thrust predictions with CFD simulations. However, their method was primarily defined for propellers of fixed-wing aircraft for which the flight condition is essentially similar to vertical flight of rotary-wing aircraft.

Challenges arise in the efficient representation of the aerodynamics of rotary-wing vehicles in forward flight, given that both thrust and inflow ratio are strongly coupled. Fundamental aerodynamic theories~\cite{leishmanbook}, namely momentum and blade element theories have been used in forward flight modeling. Momentum theory simply relates the thrust coefficient ($C_{\text{T}}$) to inflow velocity ratio by assuming the rotor as a disk through which a flux of airflow passes. Thus, it does not process any information about the blade shape.

On the other hand, the blade element theory 
uses blade geometric characteristics such as twist, chord length distribution, etc. 
Blade element theory can be used to determine the thrust coefficient using strip theory to integrate lift over the blade span. The sectional lift coefficient, $C_l$, is determined by estimating the 2D lift curve slope, $C_{l_{\alpha}}$, and the pitch and inflow angles.
\begin{equation} 
C_l=C_{l_{\alpha}} \alpha_{\text{eff}},
\label{eqn:cl}
\end{equation}
where $\alpha_{\text{eff}}=(\Theta + \alpha_{L=0} - \Phi)$ is the effective angle of attack, and $\Phi$ is inflow angle ($\lambda(r,\psi)/r$), $\lambda$ in inflow ratio, $\Theta$ is the blade pitch angle, and $\alpha_{L=0}$ is the absolute value of the zero-lift angle of attack. $\Psi$ and $r$ represent the azimuth angle and rotor radius, respectively. 

In Blade element momentum theory, the inflow ratio can be determined as a function of the flight condition parameters and geometric characteristics of the rotor. The blade momentum theory \cite{leishmanbook} was essentially developed for axial flight where both theories are rather simple and straightforward to use. The following equation shows the resultant inflow ratio: \begin{equation}
\lambda(r)=\sqrt{\left ( \frac{\sigma C_{l_{\alpha}}}{16F}-\frac{\lambda_c}{2} \right )^2+\frac{\sigma C_{l_{\alpha}}}{8F}\Theta r}-\left ( \frac{\sigma C_{l_{\alpha}}}{16F}-\frac{\lambda_c}{2} \right ),
\label{eqn:lambda}
\end{equation}
where $F$ is given as 
\begin{equation}
F=\frac{4}{\pi^2} \cos^{-1} (\exp(-f_{\text{root}})) \cos^{-1} (\exp(-f_{\text{tip}})),
\end{equation}
where $\lambda_\text{c}$ is the climb ratio, $\sigma$ is the blade solidity,
$f_{\text{root}}=\frac{N_{\text{b}}}{2}\frac{r}{(1-r)\Phi}$,
$f_{\text{tip}}=\frac{N_{\text{b}}}{2}\frac{1-r}{r\Phi}$, the function $F$ is the Prandtl's tip loss function to compensate for the loss in lift near the tip and root of the blade, with $\Theta$ being the geometric pitch, and $N_\textbf{b}$ the number of blades.
Let $-V_{\text{rel}_{B_z}}$ denote the total inlet velocity, and let $V_{\text{tip}}$ denote the blade tip velocity. In this study, $\lambda_\text{c}$ is obtained as 
\begin{equation}
    \lambda_c = -\frac{V_{\text{rel}_{B_z}}}{V_\text{tip}}.
\end{equation}
Note that $V_{\text{rel}_{B_z}}$ denotes the projection of the relative wind velocity along the positive $z$ direction of the body frame (i.e., positive inlet flow). The negative sign behind $V_{{rel}_{B_z}}$ is required, because the positive sense of $z$ is defined downward (see Fig. \ref{fig:CSs}), and a positive $\lambda_c$ implies climb for which the velocity in body frame is negative. This model can be used in forward flight.

\begin{figure}[H]
\centering
\includegraphics[width=1\textwidth]{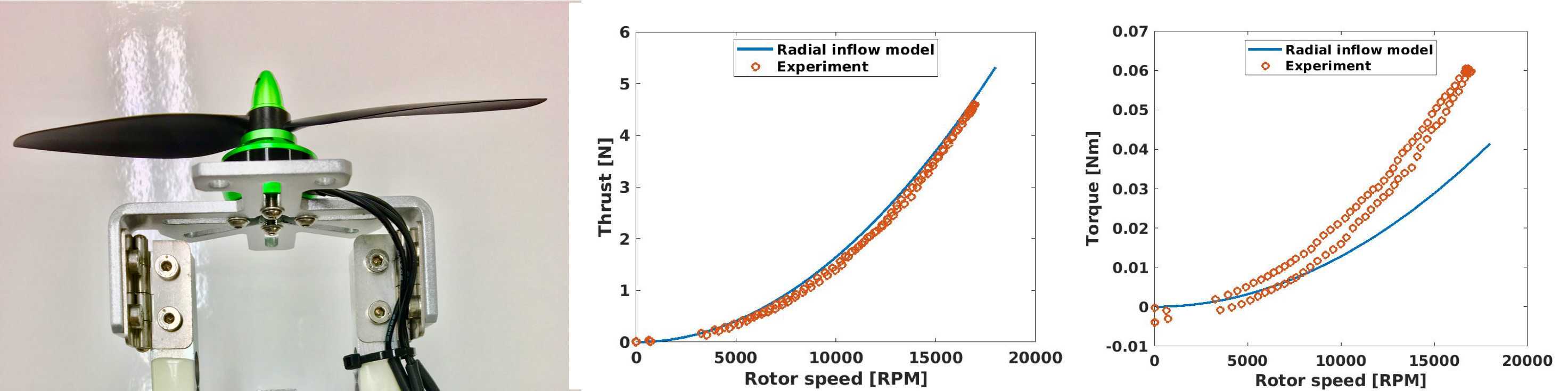}
\caption{Comparison between experimental data and the radial inflow model for hovering flight.} 
\label{fig:expvsmodel}
\end{figure}

The total thrust of a rotor is obtained by integrating the sectional lift from hub ($R_{\text{min}}=0.1R$ for the blade used in this study) to root as follows:
\begin{equation}
T=N_b \int_{R_\text{min}}^{R} \frac{1}{2} \rho C_l [(r\omega)^2 + (\lambda V_\text{tip})^2]c(r)~dr,
\label{eqn:thrust}
\end{equation}
where $c(r)$ is the chord distribution from hub to root. This model for a single rotor is compared to the experimental data of Ref.~\cite{sharma2018experimental} in Fig.~\ref{fig:expvsmodel}. The results correspond to a rotor with a radius of 7.62 cm, an average chord of 1.10 cm, and the twist distribution varying approximately from 25 to 5 degrees from root to tip. It is noted that $C_{l_{\alpha}} = 1.7059$ and $\alpha_{L=0}=4 \degree$.

As noted in Fig. \ref{fig:expvsmodel}, the thrust values predicted by the radial inflow model are in good agreement with the experimental data while the torque values are underestimated. Torque values predicted by the performance model (that will be discussed next) only represent the resisting torque due to aerodynamics, however, additional frictional resistance in the shaft of the propeller may have manifested itself in the experimental torque data.             
\subsection{Torque and Power performance model}

\begin{figure}[H]
	\centering
    \includegraphics[scale=0.35]{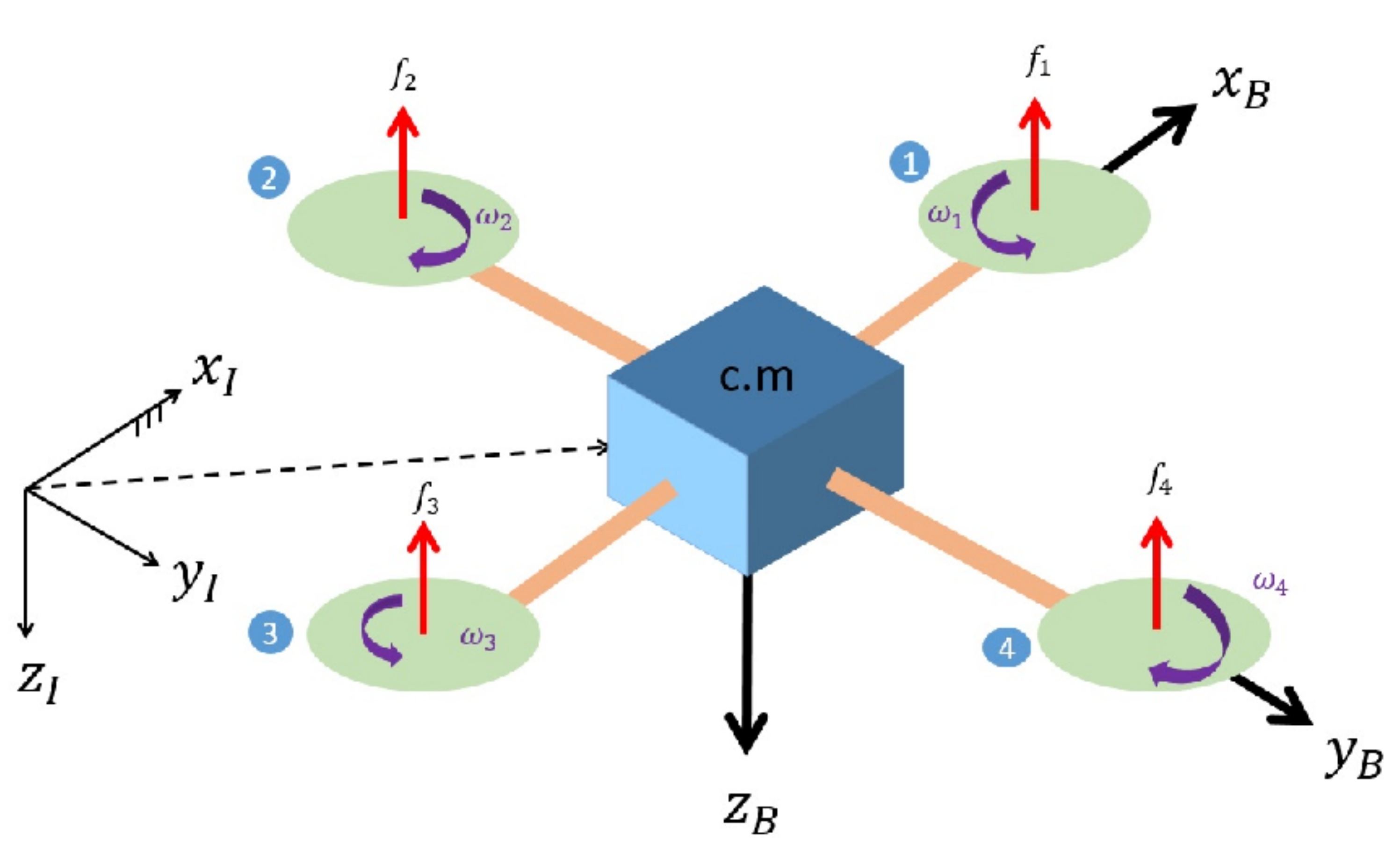}
    \caption{Definition of the body and the inertial frames of reference; positive sense of rotation and the respective positive directions of forces and torques are also presented.}
	\label{fig:CSs}
\end{figure}

For a standard quad-copter (with a `+' rotor configuration shown in Fig.~\ref{fig:CSs}), the roll and pitch moments ($\tau_\text{x}$ and $\tau_\text{y}$) are produced using differential thrust among the four rotors: 
\begin{align} \label{eqn:roll&pitch}
    \tau_\text{x} & = l (T_4 - T_2), & \tau_\text{y} & = l (T_1 - T_3),
\end{align}
where $l$ is twice the distance between the rotor and center of mass. By modulating the RPM of each rotor, it is possible to modify thrust and generate the required torque. The yaw moment, $\tau_z$ is obtained by adding the reactive yaw moment of each rotor. The reactive moment for each rotor is a function of multiple aerodynamic contributors that will be discussed in detail next.

In forward flight, the power required to overcome the resisting moment can be categorized as follows (in coefficient form, non-dimensionalized by $\rho \pi R^5 \omega^2$) :  
\begin{itemize}

\item Induced power: Power that is used to overcome lift-induced drag,  $C_{\text{P},\text{ind}}=\frac{1.15C_\text{T}^2}{2\sqrt{\lambda^2+\mu^2}}$ in which the advance ratio is defined as $\mu = \sqrt{V_{\text{rel}_{\text{B}_x}}^2+V_{\text{rel}_{\text{B}_y}}^2}/V_\text{tip}$. 

\item Blade profile power: Power required to overcome the viscous drag of each blade, $C_\text{{P,0}}=\frac{\sigma C_\text{{d0}}}{8} \left ( 1+4.6\mu^2 \right)$ where $C_\text{{d0}}$ is the profile drag coefficient.

\item Parasite power: Power required to overcome the drag exerted on the body of the vehicle due to the incoming free stream, $C_\text{{P,p}} = \frac{1}{2} \frac{f}{A} \mu^3$. It is noted that for a rotor of a quad-copter, a $1/4$ factor should be multiplied to $C_\text{{P,p}}$. $f$ is the equivalent flat plate area that models the body of the vehicle, $f/A$ can be approximated to be 0.005. 

\item Climb or descend power: the power required/produced in climbing/descending flight, $C_\text{{P,c}}=C_\text{T} \lambda_\text{c}$.
\end{itemize}
Therefore, the total power required is: 
\begin{equation} \label{eqn:power}
    C_\text{P} = \frac{1.15C_\text{T}^2}{2\sqrt{\lambda_\text{0}^2+\mu^2}}+\frac{\sigma C_\text{{d0}}}{8} (1+4.6\mu^2)+\frac{1}{8} \frac{f}{A} \mu^3+C_\text{T} \lambda_\text{c}.
\end{equation}

The power and torque coefficients are the same ($C_\text{P}=C_\text{Q}$). Thus, given Eq.~\eqref{eqn:power}, one can express the yaw torque due to the $i$-th rotor as $Q_i = C_\text{P} \rho \pi R^5 \omega_i^2$. Thus, the total reactive torque can be written as
\begin{equation} \label{eqn:taz}
\tau_\text{z}=\sum_{i=1}^4 Q_i (-1)^{i+1}. 
\end{equation}
The term $(-1)^{i+1}$ is required to make sure that the torque value associated with each rotor is taken into account with its correct sign.
\subsection{Lumped drag model}

The total drag on a quad-copter involves combinations of different aerodynamic effects. Some of the contributors were  discussed above in the context of the required torque/power to turn the four rotors at the desired speed given a flight condition. The prominent contributors to the drag of a quad-copter are induced drag, blade profile drag and translational drag due to the swirl of the induced velocity in forward flight. Parasite drag is typically small relative to the other contributors.

These parameters were studied in \cite{bangura2017aerodynamics}, and a lumped drag model for a quad-copter was introduced in Eq.~\eqref{eqn:drag} that related the drag to thrust value and relative velocity seen by the quad-copter:
\begin{equation}
\bold{D}=-\begin{bmatrix}
\bar{c}_d & 0 &0
\\ 0 & \bar{c}_d & 0
\\ 0 & 0 & 0 
\end{bmatrix}
T~\bold{V}_{\text{rel}_\text{B}},
\label{eqn:drag}
\end{equation}
where $\bar{c}_d$ is the lumped drag coefficient that was inferred from measured on-board accelerometer data for a typical quad-copter; $\bar{c}=0.04 \pm 0.0035$. In this study, $\bar{c}=0.04$ is used.

 In general, several interactions exist between the rotor blades and the wake,  and between the rotor and airframe. Diaz and Yoon \cite{ventura2018high} performed high-fidelity CFD simulations and suggested that the airframe can reduce rotor-rotor interaction, and hence increase the total thrust. With the requirement for the models to be near real-time, the  aerodynamics of each rotor is treated individually and rotor-rotor interaction effects are not considered in this study.

\section{Vehicle Dynamic Model and Control Hierarchy} \label{sec:control}

\subsection{Dynamic Modeling}
It is assumed that the vehicle consists of four rotors in a standard cross-like configuration. The schematic of the vehicle and frames of references are given in Figure \ref{fig:CSs}. The body frame, $B = \{x_\text{B},y_\text{B},z_\text{B}\}$ is needed to describe the orientation of the vehicle with respect to the inertial frame, whereas the inertial frame, $I = \{x_\text{I}, y_\text{I}, z_\text{I}\}$ is used to locate the position of the center of mass of the vehicle. The modulation of the voltage to the electrical motor of each rotor modifies the angular velocity of each propeller, $\omega_\text{i}$, $(i = 1, \cdots, 4)$, which in turn governs both the rotational and translational dynamics due to the generated forces, $f_\text{i}$, $(i = 1, \dots, 4)$. The angular velocities are paired ({\em i.e.}, the first and the third rotors rotate in a counter-clockwise manner, whereas the second and the fourth rotors rotate in a clock-wise manner) such that the net torque around the body z axis due to the rotation of propellers is zero during hovering flight.   

The translational motion of the vehicle is a consequence of the orientation (attitude) of the vehicle. In addition, it is assumed that the rotors are the only rotating components. The other components of the vehicle are assumed to have no flexibility such that rigid-body analysis can be used. It is straightforward to derive the equations of motion using the Newton-Euler method that uses the transport theorem \cite{schaub2003analytical}. 
\begin{figure}[htbp!]
	\centering
    \includegraphics[width=4in]{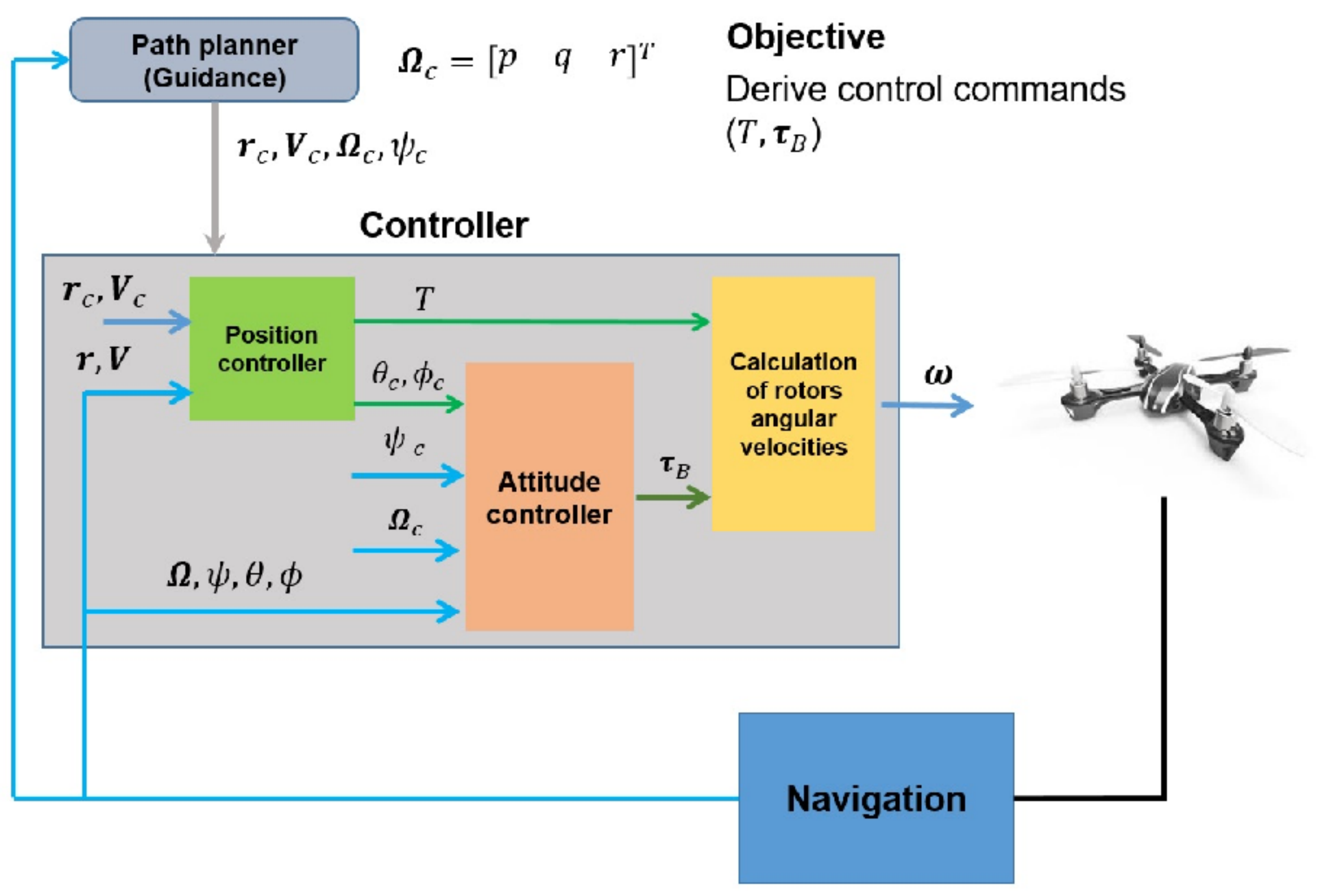}
    \caption{Outer/inner control loop scheme for position and attitude control of a quad-copter.}
	\label{fig:controlstrategy}
\end{figure}
The key point in developing an efficient control logic for quad- or multi-copters is that it is possible to achieve the overall control of the vehicle through the combination of position control and attitude control. Figure \ref{fig:controlstrategy} depicts a typical outer/inner control loop strategy along with the interconnection of the main components of an algorithm for control purposes. The outer position control loop and an inner altitude control loop. Let $\textbf{r} \in \mathbb{R}^3$ and $\textbf{v} \in \mathbb{R}^3$ denote the position and velocity vectors of the center of mass of the quad-copter, respectively, in the inertial coordinate system, and let $\bm{\Theta} = [\phi,\theta,\psi]^T$ denote the orientation angles in a standard (roll-pitch-yaw) 3-2-1 Euler rotation sequence that are used to construct the transformation matrix from the inertial frame to the body frame. Let $\bm{\Omega} = [p,q,r]^T$ denote the components of the angular velocity vector of the body frame relative to the inertial frame when expressed in the body frame - the so-called body rates. Our goal is to derive control commands, namely, thrust control $T$ and control torque $\bm{\tau}_\text{B}$ to follow a nominal trajectory. 

The nominal trajectory is usually generated in an off-line fashion to minimize a performance index (e.g, energy, usage time) of maneuver, etc. and satisfy a set of terminal constraints as well as state constraints along the trajectory. For instance, the vehicle may be constrained to keep a minimum clearance from the ground to avoid collision with a building. This task is achieved through a path-planner (or guidance algorithm) \cite{kamyar2014aircraft}. The ``Navigation'' block represents the navigation algorithm that uses sensors to measure instantaneous position, velocity and orientation of the vehicle to be used by the ``Controller'' block. 

The output of the ``Path-planner'' block is the desired time history of the states of the system that has to be followed. In Figure \ref{fig:controlstrategy}, the position, velocity, body rates and the heading angle are shown to be the outputs of the guidance block (subscript `c' is used to specify these values as commanded values that have to be tracked). These values are fed to the ``Controller'' block. At this stage, the position controller is used to track commanded values of position and velocity, i.e, $\textbf{r}_\text{c}$ and $\textbf{v}_\text{c}$. This task is achieved through a second-order differential error-tracking equation as
\begin{equation}
    \ddot{\textbf{r}}_\text{e} + \textbf{K}_\text{d} \dot{\textbf{r}}_\text{e} + \textbf{K}_\text{p} \textbf{r}_\text{e} = \textbf{0}, 
\end{equation}
where $\textbf{r}_\text{e} = \textbf{r}_\text{c} - \textbf{r}$ denotes the position error and $\textbf{K}_d$ and $\textbf{K}_\text{p}$ are coefficient matrices that are selected to be positive definite and so that to ensure acceptable time characteristics of a second-order response. A virtual control vector, $\textbf{U}$, is now defied as  
\begin{equation}
    \textbf{U} = \ddot{\textbf{r}} = \ddot{\textbf{r}}_\text{c} +  \textbf{K}_\text{d} \left (  \textbf{v}_\text{c} - \textbf{v} \right ) + \textbf{K}_\text{p} \left (  \textbf{r}_\text{c} - \textbf{r} \right ).
\end{equation}
Following  Reference \cite{zuo2010trajectory}, this virtual control input can be used along with the translational and rotational equations of motions to compute the desired thrust $T$, roll angle $\phi_c$, and pitch angle $\theta_c$. Thus, if these three variables are tracked to a good degree of accuracy, one has essentially realized the position and velocity vectors that are commanded by the path-planning algorithm. The two angles, $\theta_c$ and $\phi_c$ along with the commanded yaw altitude, $\psi_c$ and the commanded body rates $\bm{\Omega}_c$ (computed through a tracking differentiator \cite{zuo2010trajectory}) constitute the input data to the attitude controller.

The goal of the attitude controller is to track the secondary commanded values (i.e., those that are the outputs of the position controller) and those values that are commanded by the path-planning block. The output of the attitude controller is, therefore, the control torque vector, $\bm{\tau}_B$, which will result in accurate tracking of Euler angles and body rates. In the end, the thrust and torque vector are used to compute rotor angular velocity vector $\bm{\omega} = [\omega_\text{1},\omega_\text{2},\omega_\text{3}, \omega_\text{4}]^T$. The details of the algorithms can be found in Ref.~\cite{zuo2010trajectory}.

\section{Aerodynamic models and coupling to flight dynamics}
\label{sec:coupling}
In this section, the coupling of two aerodynamic models, one of which is based on Section III, to the vehicle dynamics is described. Note that the resulting model is intended for fast (real-time or near real-time) trajectory prediction and validation applications.

\subsection{Simplistic performance model}
In this approach (also known as the static model), the thrust and torque of the $i$-th rotor ($i = 1, \dots, 4$) are modeled as quadratic functions of the rotor RPM:
\begin{align} \label{eqn:QT}
Q_\text{i} & = k~ \omega_\text{i}^2, &  T_\text{i} = & b~ \omega_\text{i}^2,
\end{align}
where $k$ and $b$ are referred to as the \textit{effective} torque and thrust coefficients which can be determined experimentally, or using CFD analysis for a given rotor using simple quadratic curve fitting. For instance, this method has been used in Refs.\cite{wang2011attitude,bouadi2007modelling,fernando2013modelling,koehl2012aerodynamic}. Using experimental data shown in Fig.~\ref{fig:expvsmodel}, these coefficients can be estimated as $b= 1.5652\times 10^{-8}$ N/RPM$^2$ and $k = 2.0862\times 10^{-10}$ Nm/RPM$^2$.

Given these relations, one can form the following linear system of equations to relate the  rotor rotation rate to the required net thrust and torque. Considering Eq.~\eqref{eqn:roll&pitch}, and the fact that the sum of the thrust of each rotor is the net thrust, one can derive the following relation for the total thrust
magnitude and torque inputs: 
\begin{equation}
\begin{bmatrix}
T
\\ \tau_\text{x}
\\ \tau_\text{y}
\\ \tau_\text{z}
\end{bmatrix}
=
\begin{bmatrix} 
b & b & b & b
\\ 0 & -bl & 0 & bl
\\ -bl & 0 & 0 & bl
\\ -k & k & -k & k
\end{bmatrix}
\begin{bmatrix}
\omega_\text{1}^2
\\ \omega_\text{2}^2
\\ \omega_\text{3}^2
\\ \omega_\text{4}^2
\end{bmatrix}.
\label{eqn:lin_omg}
\end{equation}

Therefore, by solving Eq.~\eqref{eqn:lin_omg}, one can determine the rotation rate of each rotor. However, it is noted that this model is insensitive to wind conditions as well as the vehicle dynamics.

\subsection{Radial inflow model}

The radial inflow model as described in Sec.\ref{sec:aeromodel} is strictly valid for axial flight. In this work,  the incoming wind velocity is projected to the axis of rotor and the thrust is estimated using  blade element momentum theory.  The benefit of this model - in comparison to the above simplistic model -  is its sensitivity to the wind condition and vehicle dynamics. 
The inputs to this model are the required thrust, and the velocity relative to the body of the quad-copter.

Implementing this model along with the torque model introduces additional complications. First, an inverse problem should be solved, because the desired thrust is now given as an input, and RPM must be computed. At every time instant, this is performed via a simple optimization routine. Second, the yaw torque ($\tau_\text{z}$) is no longer a function of the RPM, and in fact, is a complex function as described in Eq.~\eqref{eqn:taz}. Thus, the set of equations to solve for $\omega_\text{i}$ are:
\begin{align} \label{eqn:QT2}
T &= \sum_{i=1}^4 T_i(\omega_\text{i}),   \\
\tau_\text{x} &= l (T_4(\omega_\text{4}) - T_\text{2}(\omega_\text{2})),\\
    \tau_\text{y} &= l (T_1(\omega_\text{1}) - T_\text{3}(\omega_\text{3})),\\
    \tau_\text{z} &=\sum_{i=1}^4 Q_i(\omega_\text{i}) (-1)^{i+1}.
\end{align}
It is clear that, since  $\omega_i$ are implicit  in all of the equations, an analytical solution cannot be found. Note that we need to solve for $\omega_i$ ($i = 1,\cdots,4$) at every time instant along the trajectory, which is a non-linear root-finding problem. One way to provide an approximate solution to these set of equations for every time step is as follows:

\begin{itemize}
    \item At a given time instant and a given flight condition, for every thrust $T_i$, solve the inverse problem to find $\omega_\text{i}$  and the resultant $Q_\text{i}$.
    \item For each rotor, use the relations $T_\text{i}=b_\text{i}~\omega_\text{i}^2$ and $Q_\text{i}=k_\text{i}~\omega_\text{i}^2$ to estimate the local $b_i$ and $k_i$ for that specific rotor and flight condition.
    \item use Eq.~\eqref{eqn:lin_omg}, where the coefficient matrix is given by:
    $$
    \begin{bmatrix} 
b_\text{1} & b_\text{2} & b_\text{3} & b_\text{4}
\\ 0 & -b_\text{2}l & 0 & b_\text{4}l
\\ -b_\text{1}l & 0 & 0 & b_\text{4}l
\\ -k_\text{1} & k_\text{2} & -k_\text{3} & k_\text{4}
\end{bmatrix}
    $$
    to find the final RPMs.
    \item calculate the net thrust and torque using newly found RPMs and feed them back into the dynamic model.
\end{itemize}

It has to be noted that the radial inflow model serves as a template for general aerodynamic module that can be replaced by higher-fidelity models (such as a computational fluid dynamics model).
\section{Results and Discussion}
\label{sec:results}

A quad-copter with the following physical and geometric characteristics is considered: mass, $m = 0.69$ kg, arm length, $l=0.225$ m, X-moment of inertia, $I_x= 0.0469$ kg m$^2$, Y-moment of inertia, $I_y= 0.0358  $ kg m$^2$, Z-moment of inertia, $I_z=  0.0673$ kg m$^2$, rotor moment of inertia, $I_r = 3.357 \times 10^{-5} $ kg m$^2$.

The aerodynamic models for the quad-copter, and the wind models (Section II) were integrated with the control module in the Simulink environment of MATLAB. 
We have performed flight simulations for two representative nominal trajectories: 1) an ascent-straight-descent trajectory, 2) a circular trajectory.

\subsection{Ascent-straight-descent path}
\begin{figure}
\centering
\includegraphics[width=3.0in]{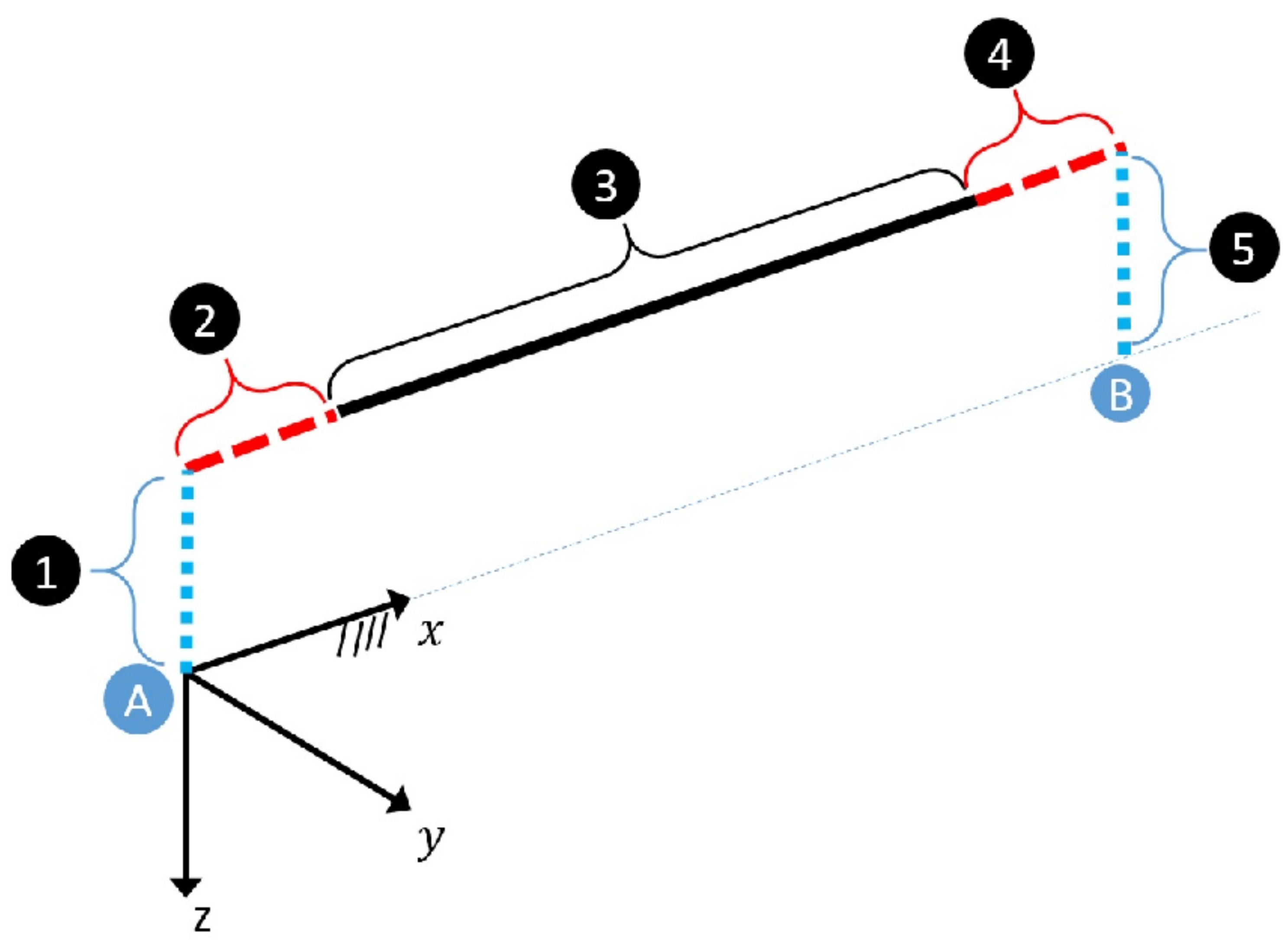}
\caption{Schematic of the ascent-straight-descent nominal trajectory.}
\label{fig:ADC_details}
\end{figure}

Let $\Delta t_i$ denote the time interval of the $i$-th segment of a multi-segment trajectory. Figure \ref{fig:ADC_details} shows the schematic of an idealized rectangular path that consists of five segments: 1) taking off vertically to an altitude of 40 m (where the initial and final vertical velocities are zero) over a time interval of $\Delta t_1 = 10$ seconds, 2) accelerating from zero forward velocity to a cruise constant speed of 15 m/s over a time interval of $\Delta t_2 = 12$ seconds, 3) continuing with the cruise speed for $\Delta  t_3 = 30$ seconds, 4) decelerating from a forward velocity of 15 m/s to zero over a time interval of $\Delta t_4 = 15$ seconds, and 5) descending over a time interval of $\Delta t_5 = 10$ seconds. As it follows the trajectory, the quad-copter is subject to the wind field described in Section \ref{sec:wind}. For all segments (except for the cruise segment \#3), a cubic polynomial is used to enforce the boundary conditions on $x$, $y$ and $z$ position and velocity coordinates \cite{taheri2012shape}. For instance, there are four boundary conditions on the translational velocity of the first segment (\textit{i.e.}, $x_i$, $x_f$, $\dot{x}_i$, and $\dot{x}_f$) where subscripts `i', and `f' denote the initial and final times of the segment. It is possible to fit a cubic polynomial using the prescribed boundary conditions and solve for the coefficients of the cubic polynominal in terms of the boundary conditions. Velocity and acceleration data of the nominal trajectory are known by taking the first and second time derivatives of the position coordinate. It is noted that the planned trajectory information consists of the planned position, velocities and acceleration, as well as the heading (yaw) angle ($\psi(t) = 0$) all of which are used in the attitude controller block in Fig.\ref{fig:controlstrategy}. The results for different flight parameters are next presented using a full-wind representation and radial inflow model. 

\begin{figure}[H]
\centering
\includegraphics[width=3.0in]{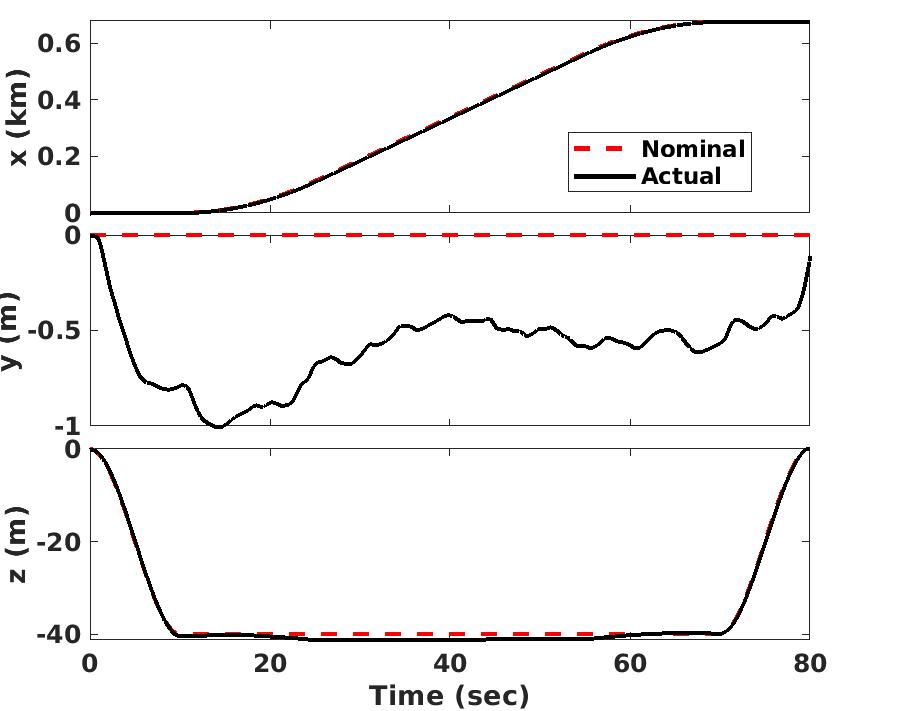}
\caption{Time histories of the inertial coordinates of the trajectory of the quad-copter compared to the nominal trajectory.}
\label{fig:location_straight}
\end{figure}

\begin{figure}[H]
\centering
\includegraphics[scale=0.25]{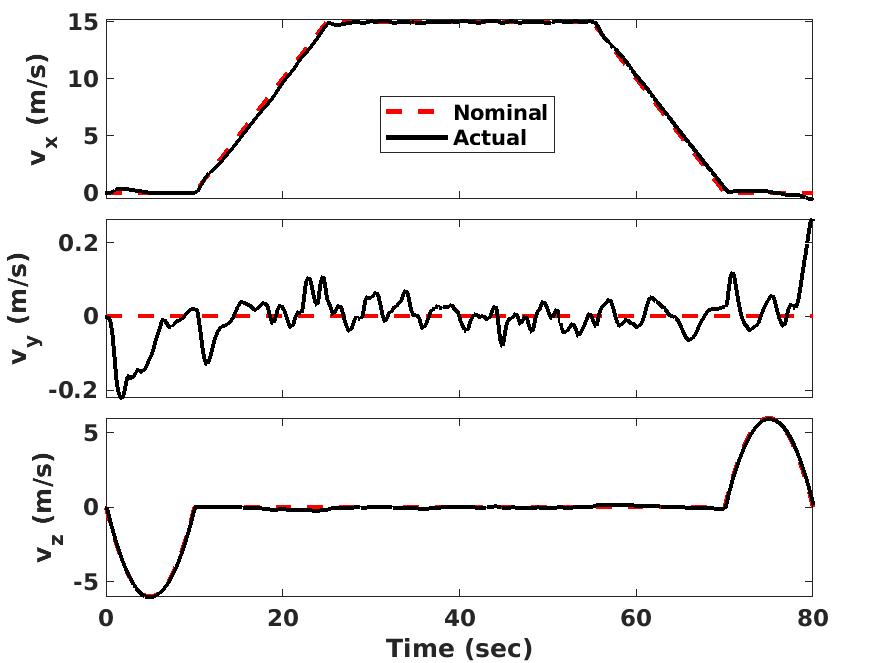}
\caption{Time histories of the inertial velocity component of the trajectory of the quad-copter vs. the nominal velocity.}
\label{fig:vel_straight}
\end{figure}

\begin{figure}[H]
\centering
\includegraphics[scale=0.25]{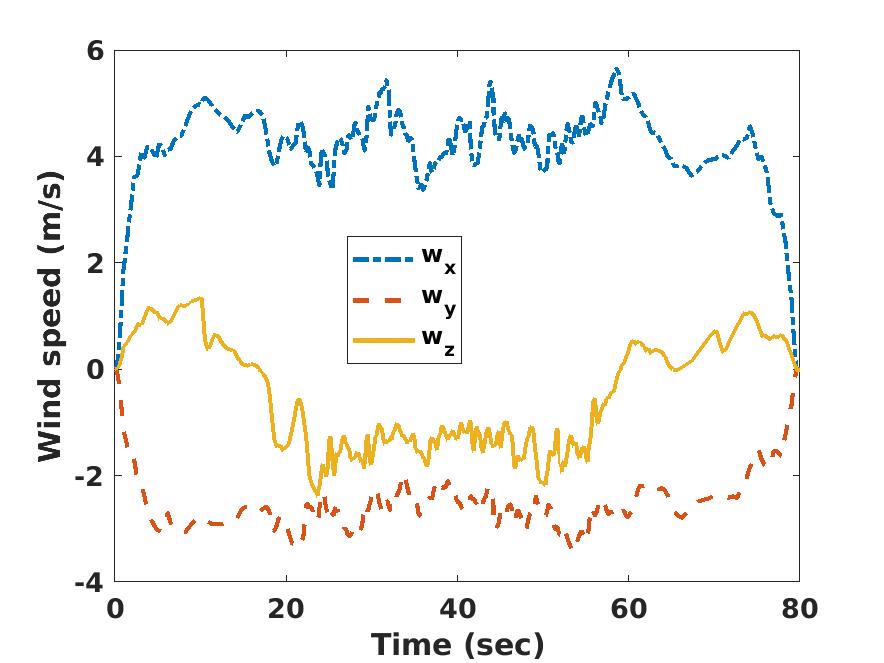}
\caption{Components of the wind velocity along the actual trajectory expressed in the inertial frame.}
\label{fig:windspeed_straight}
\end{figure}

The results demonstrate that the planned trajectory and the vehicle attitude and position are controlled successfully. The time histories of the quad-copter position and velocity coordinates as well as its attitude are shown in Figures~\ref{fig:location_straight}-\ref{fig:windspeed_straight}. All of the resultant flight parameters (shown in color red dashed line) are compared with their planned ones (shown in  black sold line). The resultant position versus planned position is shown in Fig.~\ref{fig:location_straight}.  
As noted, the quad-copter has tracked the planned trajectory with acceptable accuracy. The off-the-track, $y$ position coordinate appears to have the largest deviation due to the side wind effects. The positive $x$ direction of the considered inertial frame points to the east (that is geographic north is in the direction of ($-y$)). In the considered wind model, the wind blows from south west (i.e., from $-x$ to $+x$ and $+y$ to $-y$), and that would push the quad-copter to its left side ($-y$) and to forward ($+x$). The planned and resultant velocities in the inertial body frame are shown in Fig.~\ref{fig:windspeed_straight}, which reveals the fluctuations due to turbulent gusts.


The wind velocity at the vehicle CG is shown in Fig. \ref{fig:windspeed_straight}, illustrating a desirable forward wind and an undesirable side wind experienced by the vehicle. 

\begin{figure}[H]
\centering
\includegraphics[scale=0.20]{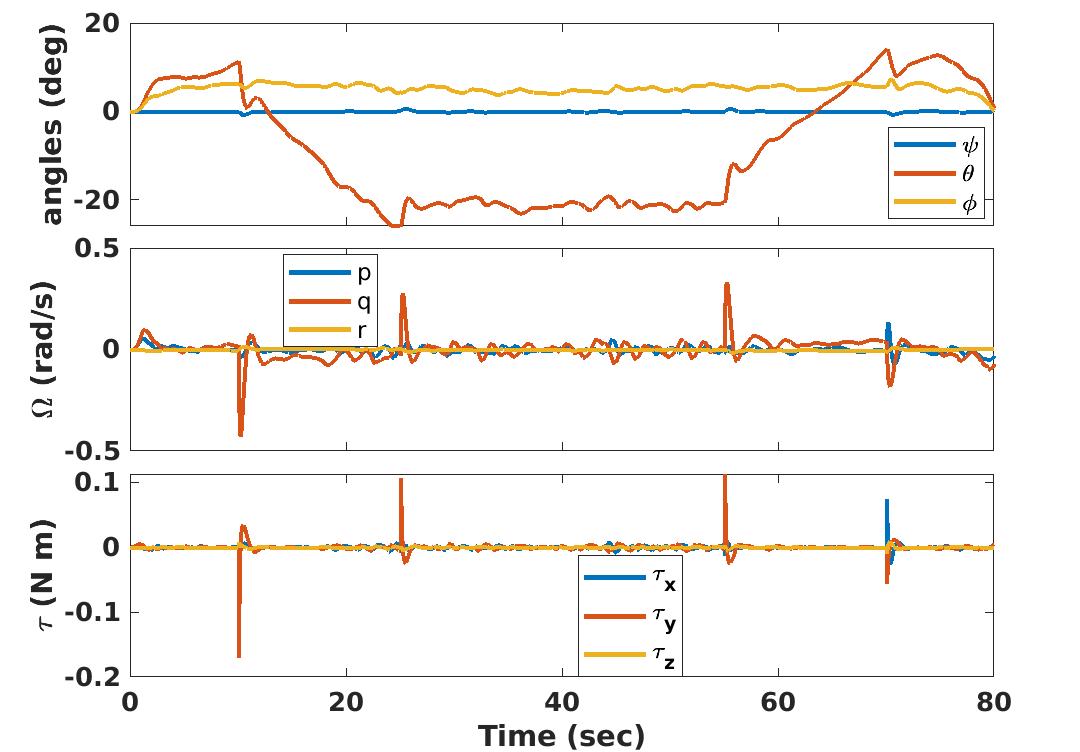}
\caption{Time histories of the Euler angles, body rates and control torques during flight.}
\label{fig:euler_straight}
\end{figure}

\begin{figure}[H]
\centering
\includegraphics[scale=0.24]{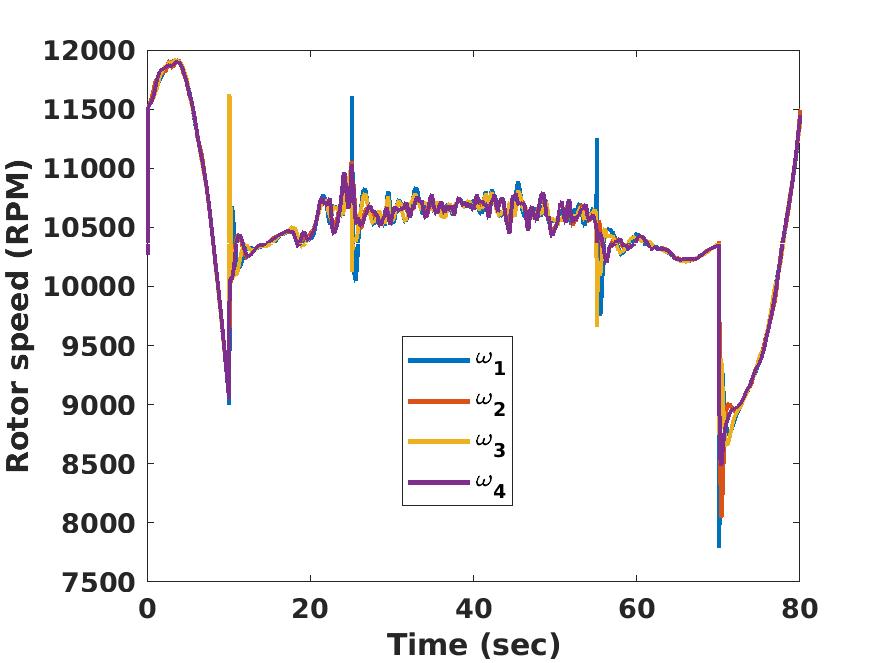}
\caption{Time histories of the rotors' RPM along the trajectory.}
\label{fig:omegas}
\end{figure}

The required rotor RPMs to track the planned path are obtained using the radial inflow model and  shown in Fig. \ref{fig:omegas}.

\begin{figure}[H]
\centering
\includegraphics[scale=0.25]{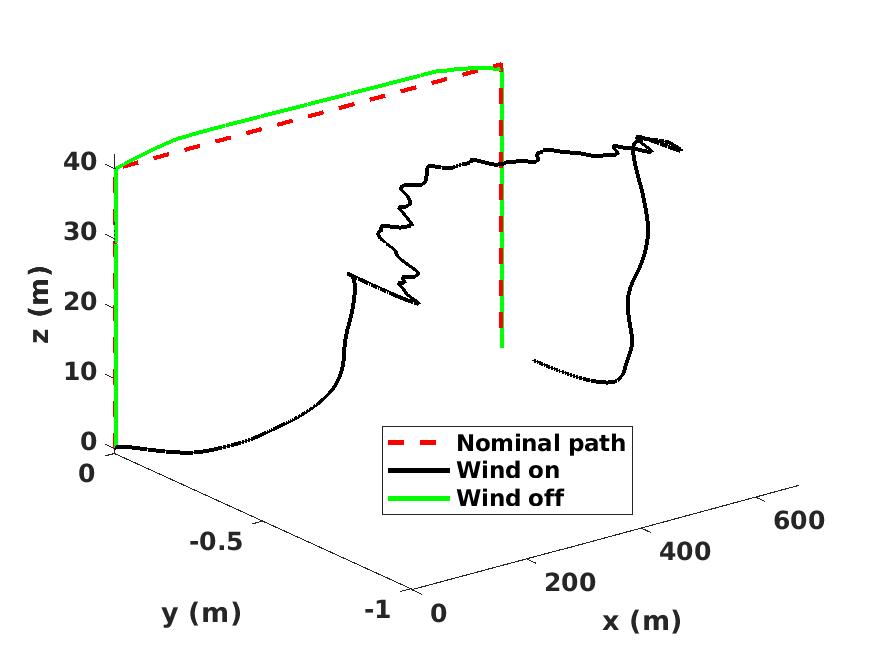}
\caption{Flown trajectories with and without consideration of wind (each axis is scaled differently for clarity).}
\label{fig:windpathyn}
\end{figure}

\begin{figure}[H]
\centering
\includegraphics[scale=0.25]{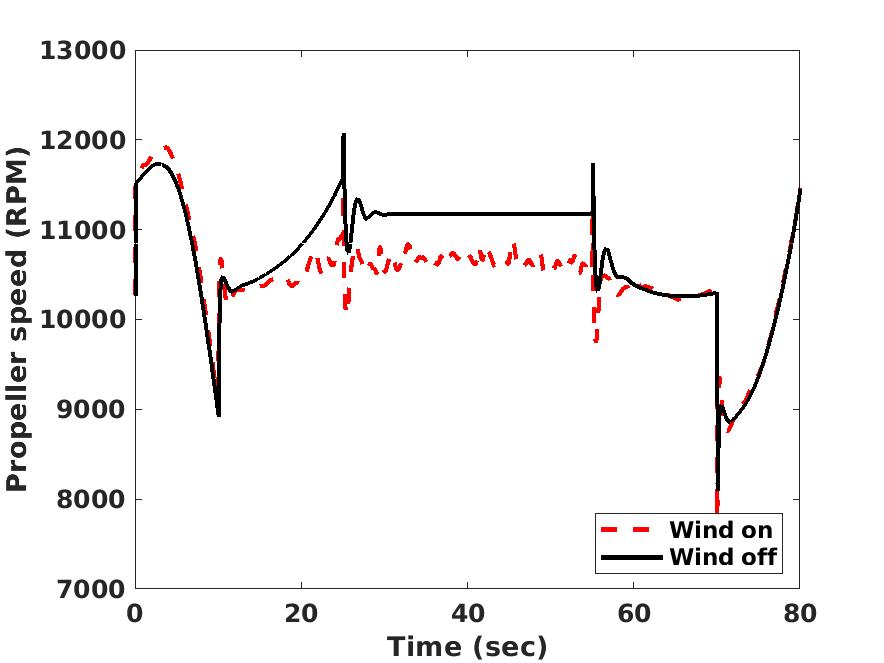}
\caption{Comparison of the rotor \#1 RPM with and without consideration of wind.}
\label{fig:rpmsdiff}
\end{figure}

The effect of the wind on the resultant trajectory and vehicle dynamics are shown in Figures~\ref{fig:windpathyn} and \ref{fig:rpmsdiff}. It is noted that without a wind field, the planned path was tracked  with almost no deviations. It is also evident that wind effects on the RPM inputs are more prominent in the cruise section compared to the vertical take-off and landing segments. 

\begin{figure}[H]
\centering
\includegraphics[scale=0.25]{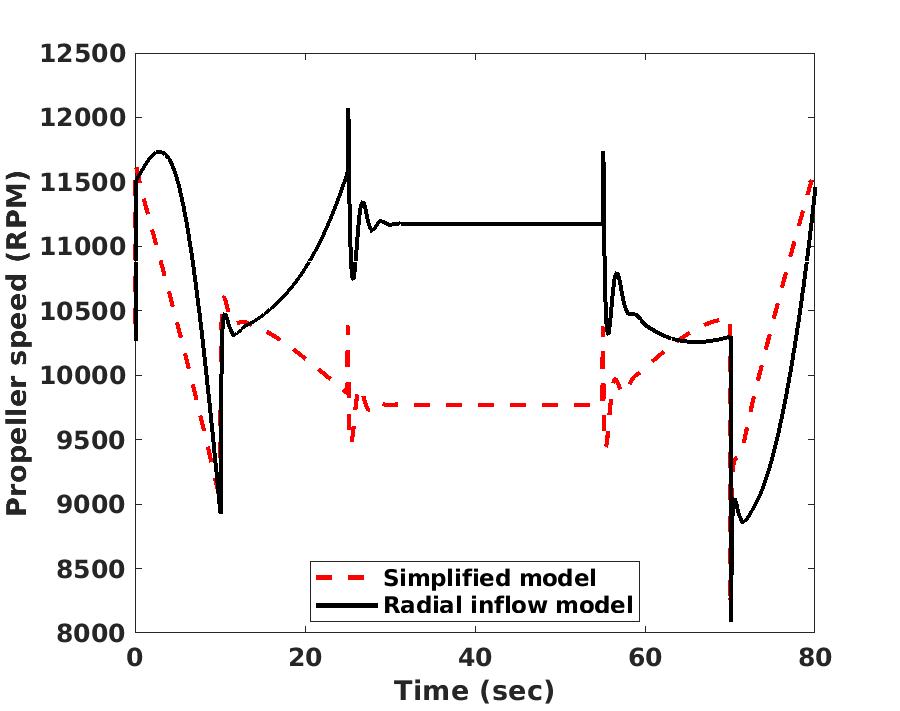}
\caption{Rotor RPM prediction with different models.}
\label{fig:diffmod}
\end{figure}

The difference between the simplified and radial inflow models in terms of the predicted rotor speeds is compared in Fig.~\ref{fig:diffmod}. Note that the resultant RPM of only rotor \#1 (leading rotor, see Fig. \ref{fig:CSs}) in a no wind condition is presented for clarity in  depicting the discrepancy. The predicted RPMs from both models are very similar in axial flight for the hover case as shown in Fig. \ref{fig:expvsmodel} ($t < 10$ s and $t > 70$ s). There is, however, a large discrepancy in the cruise section of the trajectory, ($10<t<70$ s) where the velocity relative to the body of the quad-copter increases the inflow  $\lambda$ (see Eq.~\eqref{eqn:lambda}). Therefore, the lift coefficient (Eq.~\eqref{eqn:cl}) decreases, because $\Phi$ increases. Hence, the rotor speed has to increase to maintain the required thrust.

The torque and power performance model, introduced in Section \ref{sec:aeromodel}.A, can be used to provide an estimate of the total required power during flight. Considering Eq.~\eqref{eqn:power}, one can simply write the total power as:
\begin{equation}
    P=\sum_{i=1}^4 C_{\text{P}_i} \rho \pi R^5 \omega_i^3. 
\label{eqn:totat_power}    
\end{equation}

\begin{figure}[H]
\centering
\includegraphics[scale=0.25]{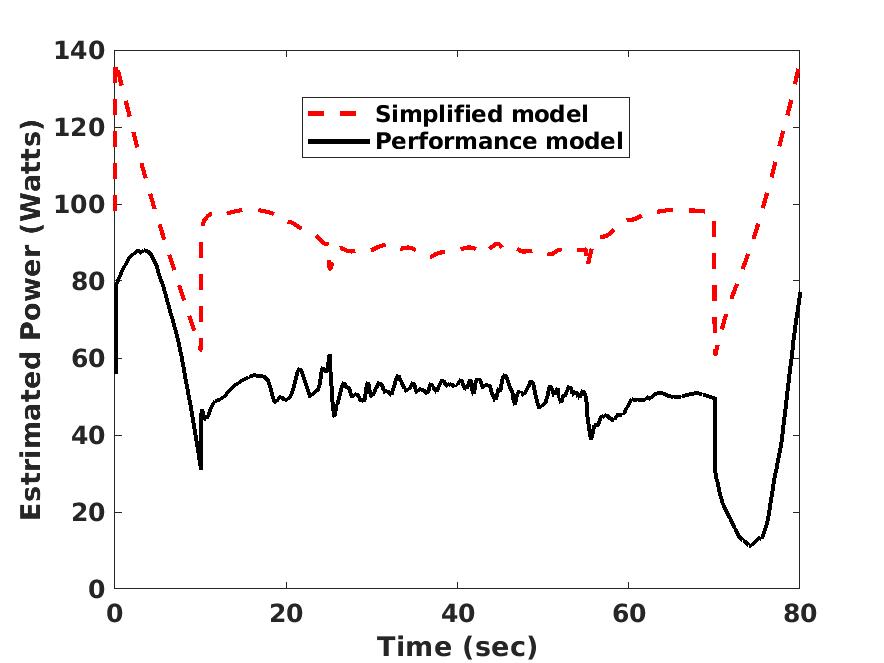}
\caption{Time histories of the required power estimates using different models.}
\label{fig:power}
\end{figure}

For the simplified model, the power estimate is simply $\sum_{i=1}^4 k\omega_i^3$ where the value of $k$ is obtained from the torque data of Fig. \ref{fig:expvsmodel}. The time history of the  power estimated by both models is shown in Fig. \ref{fig:power}.

The simplified model yields larger power values and this is mainly due to the offset seen between the torque predictions of performance model and experimental data in Fig. \ref{fig:expvsmodel}. It is also apparent that take-off ($t<10$ s) requires more power  than landing ($t>70$ s). The radial inflow model represents this effect, while the simplified model is unable to do this. 

\begin{figure}[H]
\centering
\includegraphics[scale=0.25]{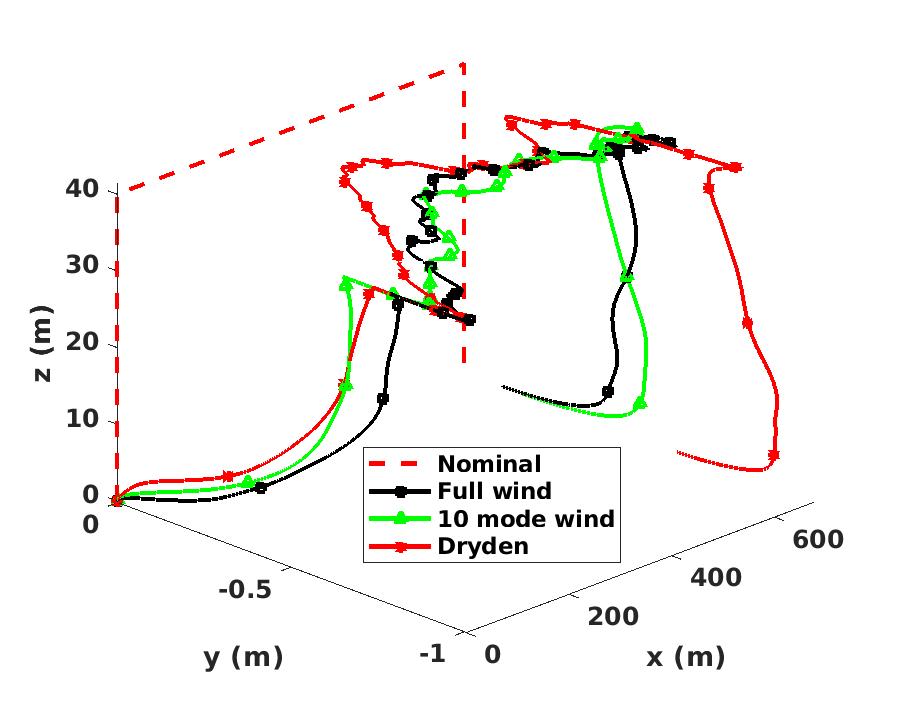}
\caption{Effect of different versions of wind model on the resultant trajectory.}
\label{fig:diffwind}
\end{figure}

To assess the importance of the wind model, the wind field is reconstructed using different number of modes, and flight simulations are performed to illustrate the impact on the results.  A comparison between the resultant trajectories of the full wind, reduced-order versions of the wind field, and the Dryden Wind Turbulence model is shown in Fig. \ref{fig:diffwind}. In these simulations, the radial inflow  model was used as the aerodynamic model to compare the impact of the different wind models. While the controller attempts to keep the quad-copter on the nominal path regardless of the  wind model used, it is clear that the Dryden model - as a consequence of the fact that the fluctuations are not spatio-temporally correlated - results  in a lesser deviation from the nominal trajectory. 

\begin{figure}[H]
\centering
\includegraphics[scale=0.25]{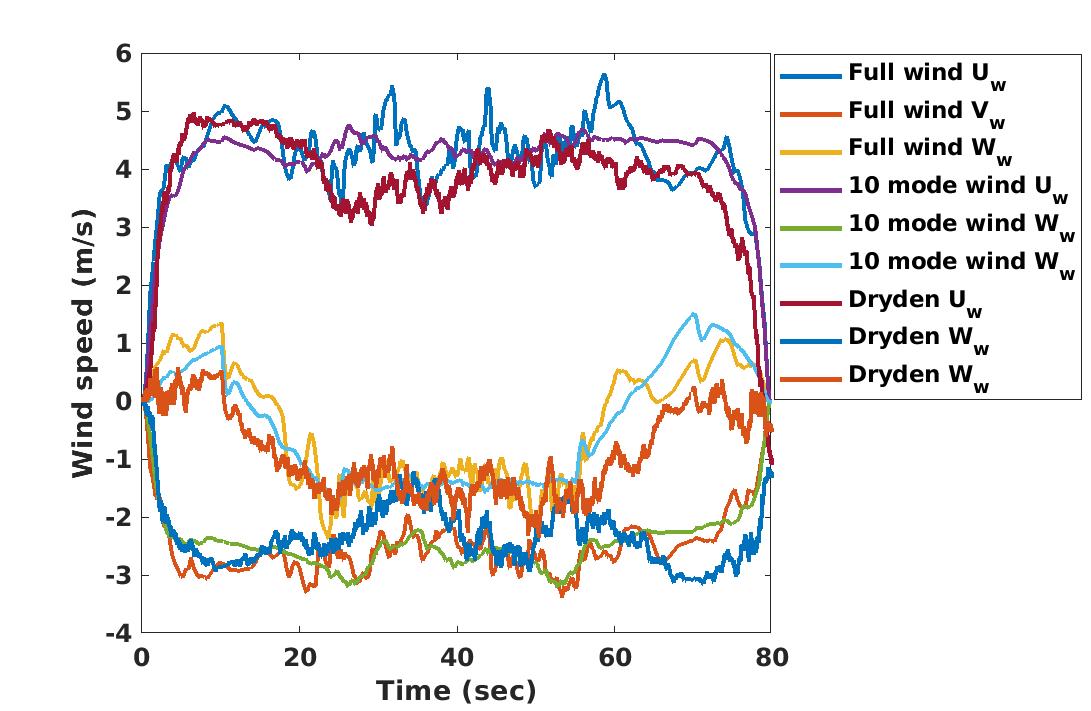}
\caption{Wind velocity components in the inertial frame.}
\label{fig:fullwind}
\end{figure}

The incoming velocities in the three directions seen by the quad-copter for different wind models are shown as a function of time in Fig.~\ref{fig:fullwind}.

\subsection{Circular path}
\begin{figure}[H]
\centering
\includegraphics[scale=0.15]{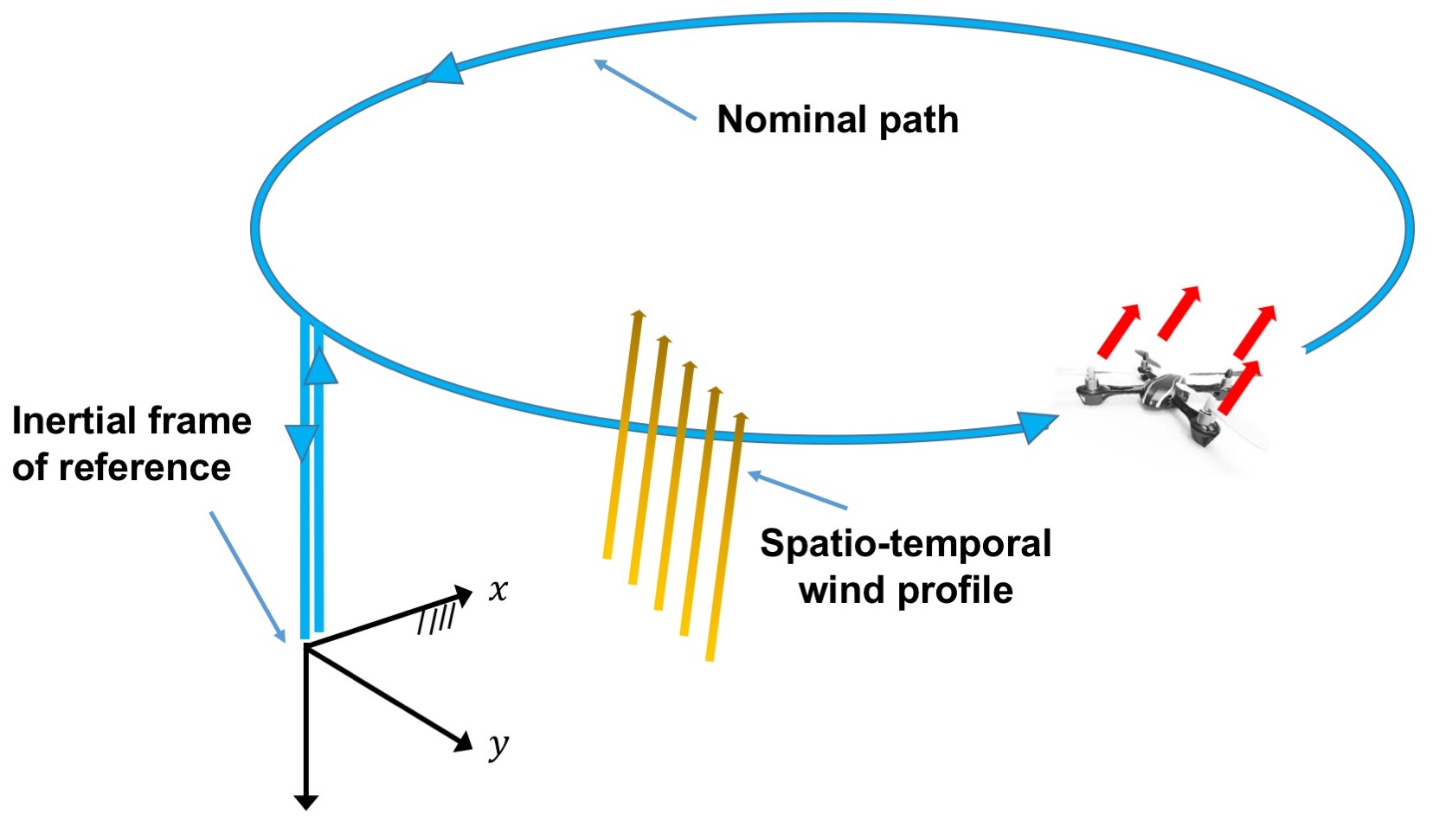}
\caption{Schematic of the circular path.}
\label{fig:sch_circ}
\end{figure}

Flying in a circular path further accentuates the importance of the flight controller and wind model. In this case, the quad-copter has to follow a path with continuous acceleration. The presence of wind has both favourable and unfavourable effects during portions of the trajectory, as  discussed herein. A schematic of the circular path and the wind condition is shown in Fig.\ref{fig:sch_circ}.
Similar to the straight path, the trajectory for the circular path consists of five segments: 1) taking off to an altitude of 60 m, 2) accelerating for 10 seconds to a nominal speed while being in the circular path with radius of 80 m, 3) following the circular path with a nominal speed, 4) decelerating for 10 seconds to stop at the point where the circular path was initially started, and 5) landing.

\begin{figure}
\centering
\includegraphics[scale=0.25]{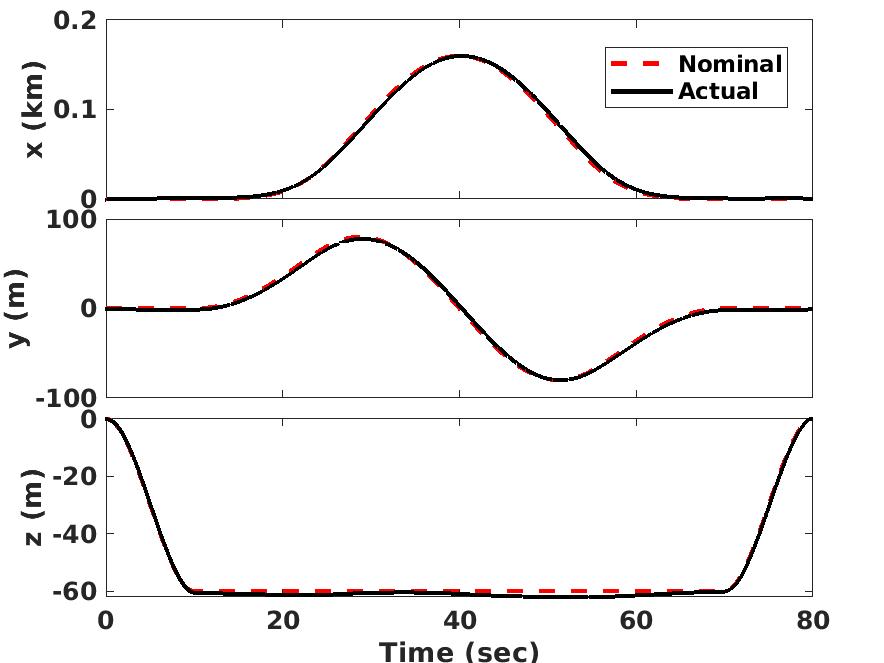}
\caption{Time histories of the position coordinates of the center of mass of the quad-copter for circular path.}
\label{fig:location_circ}
\end{figure}

\begin{figure}
\centering
\includegraphics[scale=0.25]{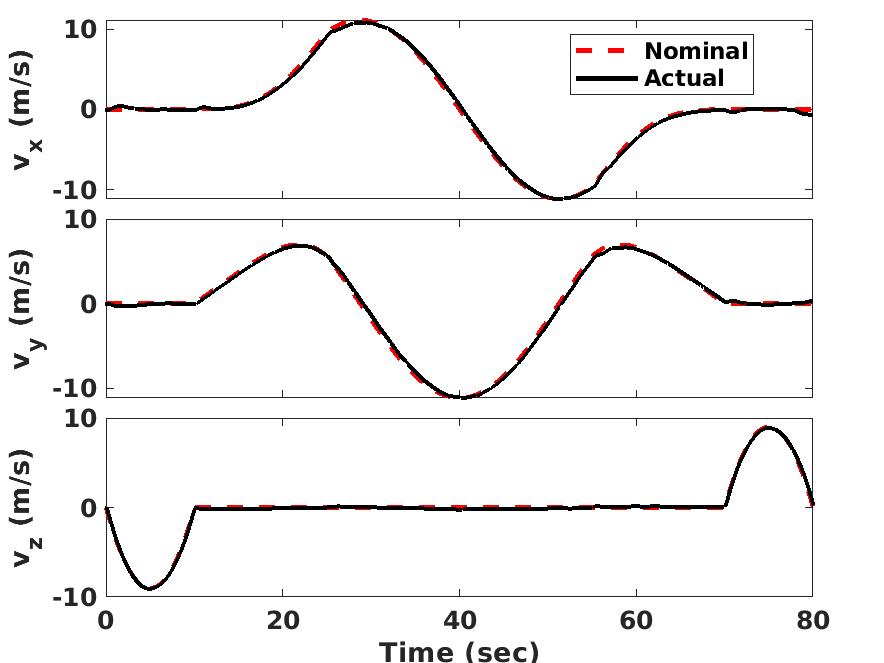}
\caption{Time histories of velocity coordinates of the center of mass of the quad-copter for circular path.}
\label{fig:vel_circ}
\end{figure}

A comparison between the planned and obtained location and velocity in the three directions is shown in Figs.~\ref{fig:location_circ} and \ref{fig:vel_circ}. The results are indicative of the fact the vehicle has been able to track the planned path well and maintain the quad-copter on the nominal path. 

\begin{figure}
\centering
\includegraphics[scale=0.25]{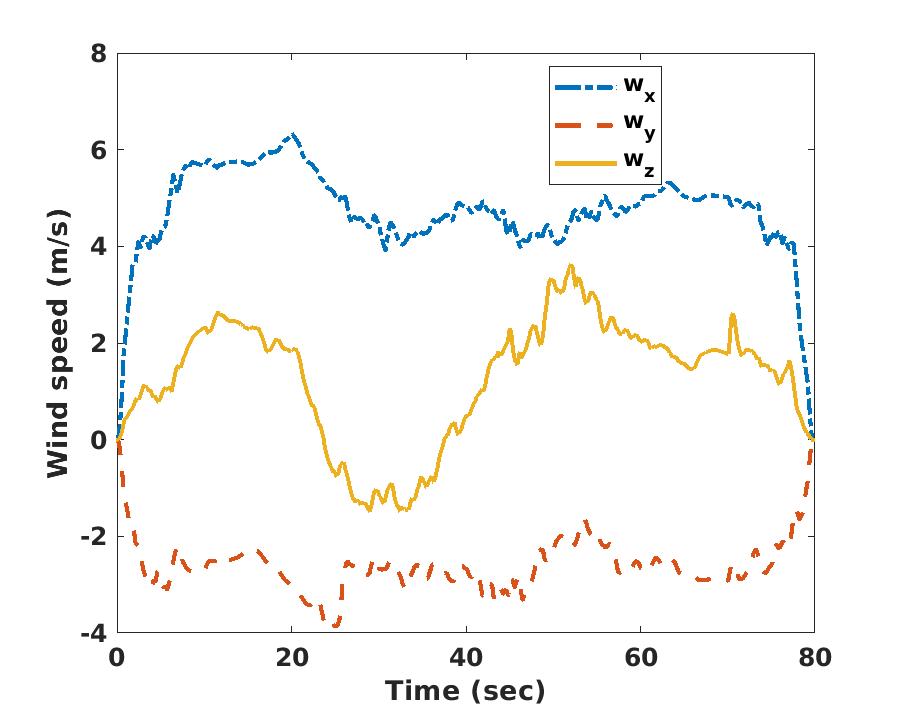}
\caption{The wind velocity at the center of mass of the quad-copter.}
\label{fig:windspeed_circ}
\end{figure}

The wind velocity at the center of mass of the quad-copter is shown in Fig. \ref{fig:windspeed_circ}. The values are slightly larger compared to the straight path (see Fig. \ref{fig:windspeed_straight}) given that the circular path has a higher altitude.

\begin{figure}
\centering
\includegraphics[scale=0.20]{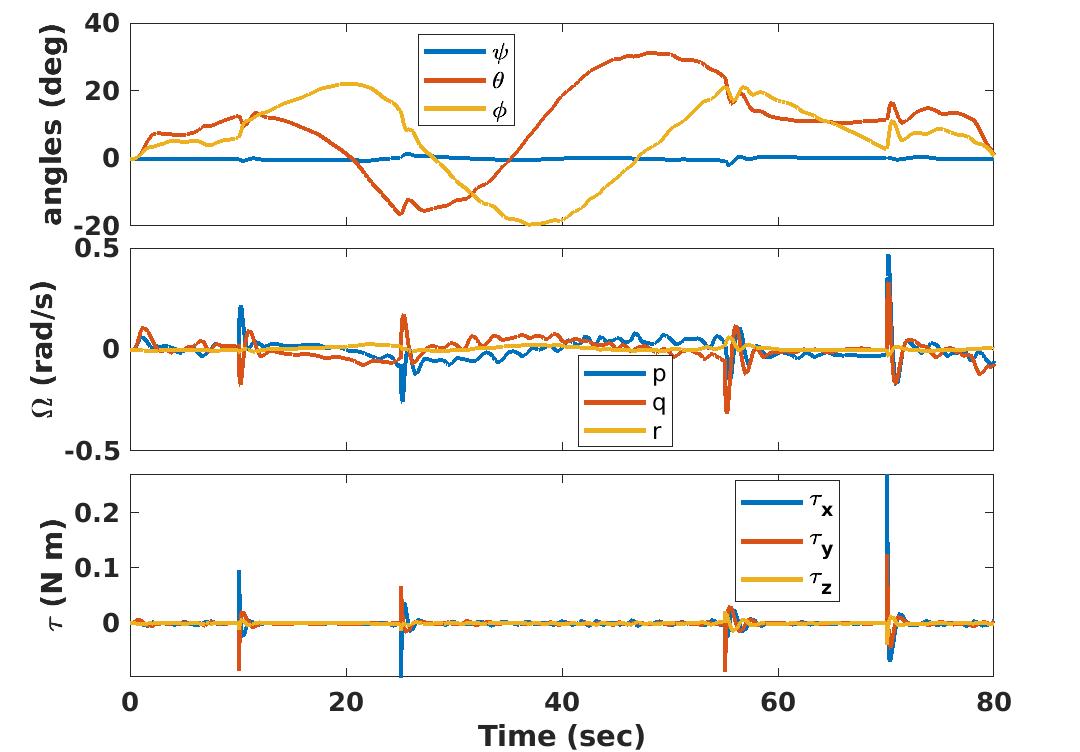}
\caption{Euler angles (roll, pitch and yaw), body rates, and torques applied to the quad-copter during circular path.}
\label{fig:euler_circ}
\end{figure}

For this path, the commanded heading angle ($\psi_c$) was set to zero which means the quad-copter does not turn around its Z-axis, and the circular path was tracked by controlling only roll and pitch angles. The Euler angles, and their rated and resultant torques are depicted in Fig. \ref{fig:euler_circ}.

\begin{figure}
\centering
\includegraphics[scale=0.25]{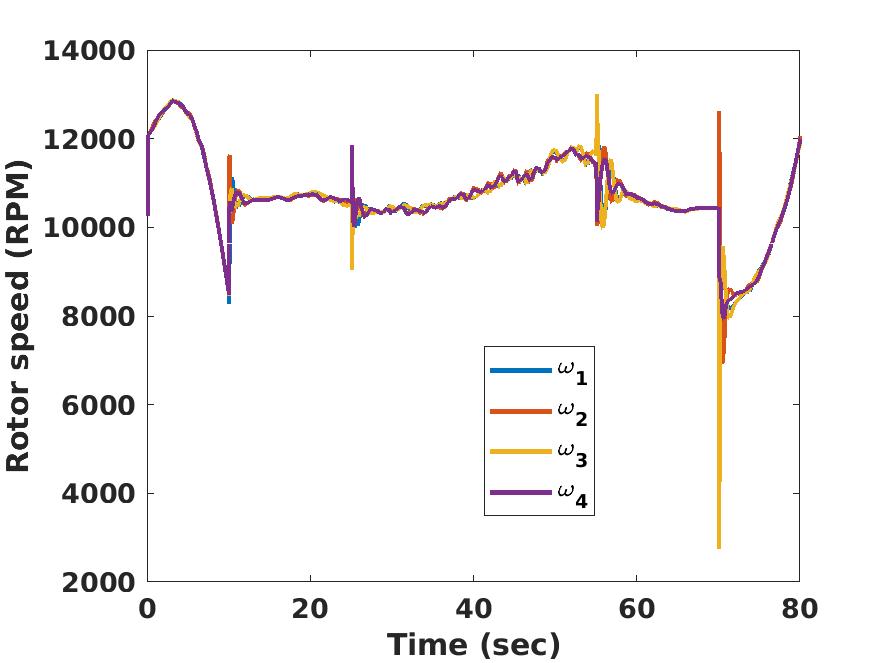}
\caption{Rotor speeds to maintain the quad-copter on the circular path.}
\label{fig:omegas_circ}
\end{figure}

As noted from Fig.~\ref{fig:euler_circ}, for the take-off phase ($t<10$ s), the quad-copter has to roll (right rotor (\#3) down), and pitch up (front rotor (\#1) up) to negate  the incoming wind from the south west, and then continue the entire path with a nose up position to negate the effect of the wind. The corresponding four rotor speeds shown in Fig.~\ref{fig:omegas_circ} were obtained as the controller commands computed to maintain the planned circular trajectory.

\begin{figure}
\centering
\includegraphics[scale=0.25]{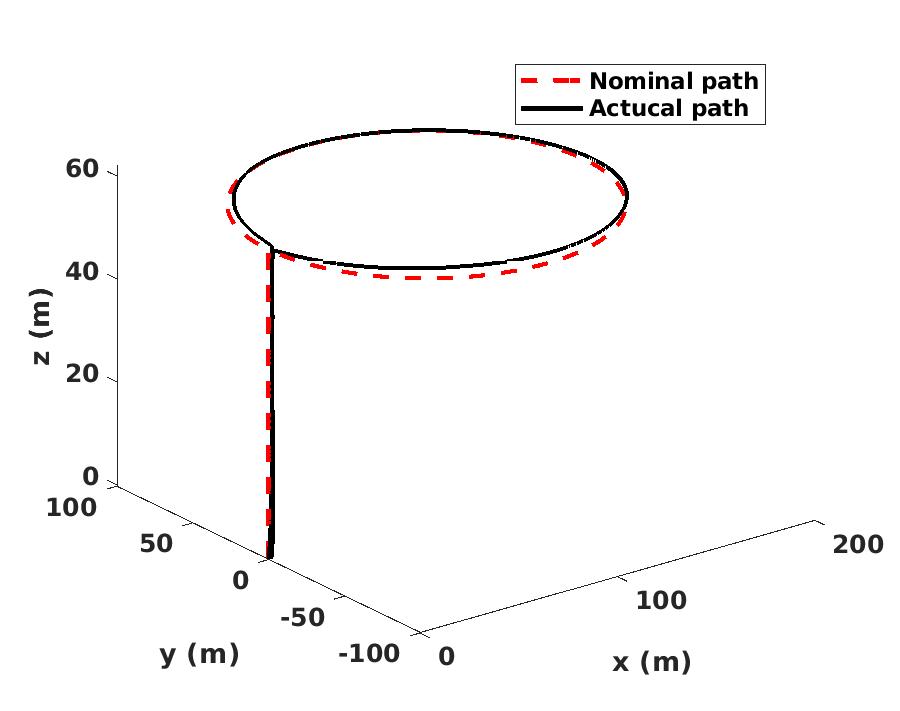}
\caption{Trajectories with and without consideration of wind for the circular path.}
\label{fig:3D_circ_pos}
\end{figure}

\begin{figure}
\centering
\includegraphics[scale=0.25]{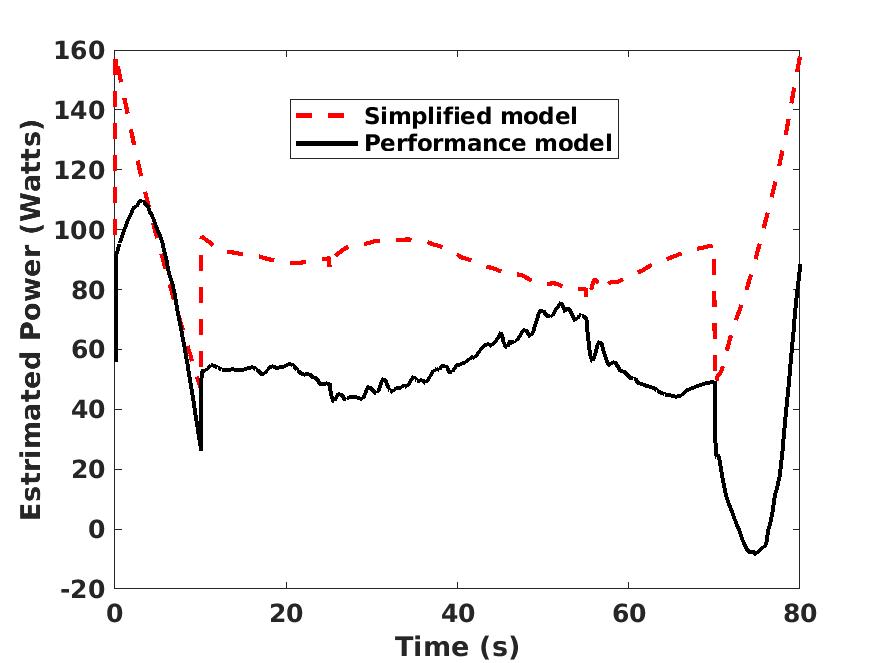}
\caption{Estimated required power through different models for circular path.}
\label{fig:power_circ}
\end{figure}

The actual versus the planned trajectories are demonstrated for the circular path in Fig.~\ref{fig:3D_circ_pos}. As expected, the actual trajectory is shifted toward positive $x$ and negative $y$ due to the wind condition. A maximum deviation of 2 meters between the two paths is noted. The estimates of the required power obtained from the simplified and radial inflow models are shown in Fig. \ref{fig:power_circ}.

When the  quad-copter is on the circular trajectory ($10<t<70$ s), the simplified model shows a higher sensitivity to the favorable and adverse wind conditions. In the first half circle ($10<t<40$ s), the wind is overall favorable. In the second half circle ($40<t<70$ s), the wind causes unfavorable drag as well as more induced inflow to the rotors that leads to higher RPMs (see Fig. \ref{fig:omegas_circ}), and subsequently, higher power as depicted in Fig. \ref{fig:power_circ}.

\subsection{Optimal cruise speed}

\begin{figure}
\centering
\includegraphics[scale=0.23]{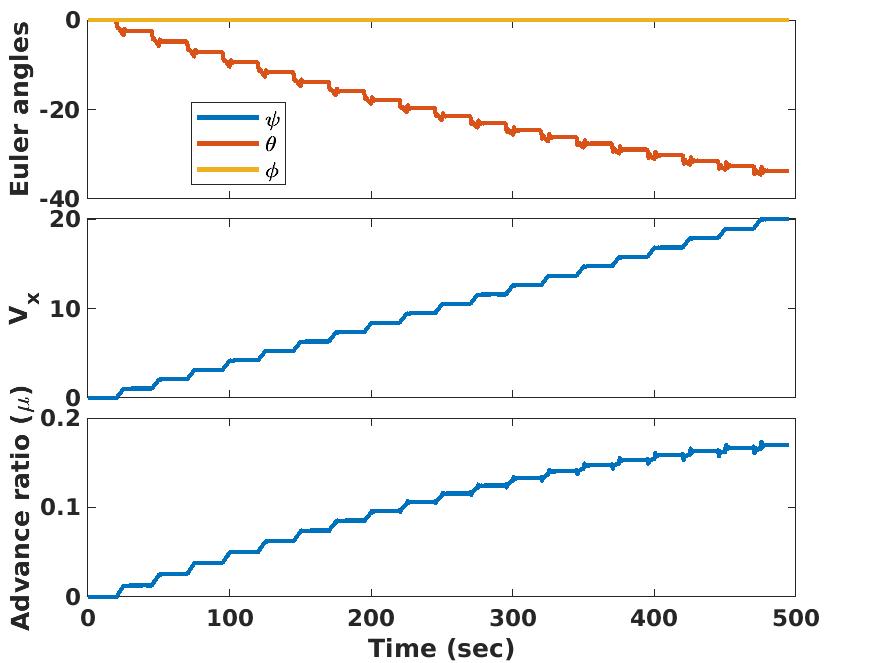}
\caption{Euler angles and cruise speed for optimal cruise speed determination.}
\label{fig:opt_angle_speed}
\end{figure}

The relationship between  power and cruise speed is non-monotone and non-linear. The power model (see Eq.~\eqref{eqn:power}) can be further analyzed to determine the optimal cruise speed and compared with the power required in hover.  A new trajectory was defined, in which the quad-copter starts from hover, and the forward speed is adjusted  incrementally on a straight path. Within each increment, it accelerates for $5$ seconds to add $1$ m/s to its speed during the acceleration phase, and stays on that specific cruise speed for $20$ seconds. This increment is performed $20$ times, and the vehicle will eventually reach a forward speed of $20$ m/s in a total time of 500 seconds. The Euler angles and forward velocity in the inertial reference frame, and advance ratio are shown in Fig. \ref{fig:opt_angle_speed}.

\begin{figure}
\centering
\includegraphics[scale=0.28]{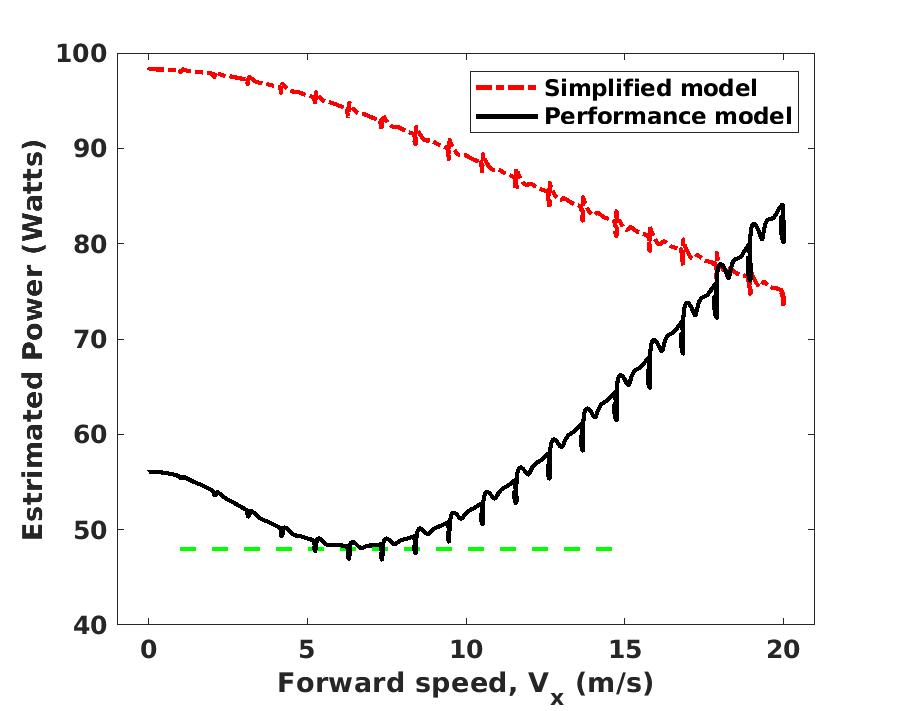}
\caption{Estimated required power versus cruise speed for simplified and radial inflow models.}
\label{fig:opt_power}
\end{figure}

It is noted that the advance ratio, $\mu$, is relatively small for the entire flight. At the highest forward speed of $20$ m/s, the advance ratio is $\mu=0.17$. The pitch angle increases in magnitude during the acceleration phase and maintains the same level during the constant cruise speed part of each increment. The speed of the quad-copter as a function of the leading rotor \#1 (see Fig. \ref{fig:CSs}) rotational speed is shown in Fig. \ref{fig:opt_power}.

\begin{figure}
\centering
\includegraphics[scale=0.27]{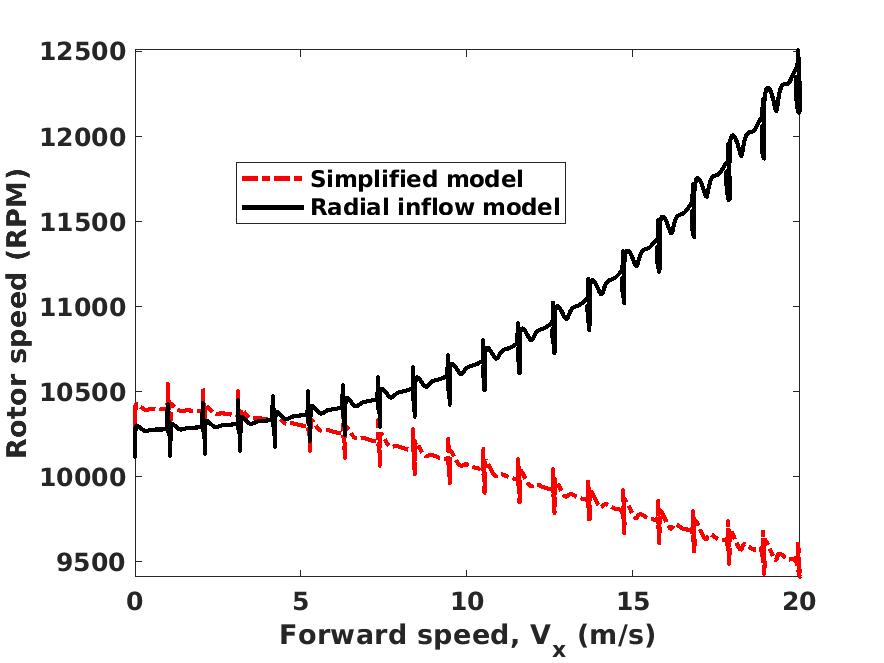}
\caption{Estimated rotor \#1 speed as a function of cruise speed.}
\label{fig:opt_rpm}
\end{figure}

As noted from Fig. \ref{fig:opt_power}, the power curve with the performance model (see Eq. \ref{eqn:power}) has a local minimum around $V_\text{x}=7.2$ m/s (indicated with the green dashed line). It is also noted that the simplified model that only uses the rotor speed to estimate the power ($P= k \sum_{i=1}^4\omega_i ^3$) estimates less power consumption for higher speeds. This is due to the fact that at higher speeds, the drag (see Eq.~\eqref{eqn:drag}) provides an upward component when the pitch angle is negative ($\theta<0$), thus augmenting thrust. Therefore, the thrust required by the rotor (to sustain the altitude) decreases and based on the simplified model, the required RPM should also decrease. The estimated RPM values of rotor \#1 versus forward speed from both models are shown in Fig. \ref{fig:opt_rpm}.

As expected, the two models show opposite trends. Based on Figs. \ref{fig:opt_power} and \ref{fig:opt_rpm}, higher RPM does not necessarily indicate higher required power, as the flight condition has a significant impact on the total power.







\section{Summary and Conclusion} \label{sec:sum}
A comprehensive suite of tools was introduced for performing realistic flight simulations for unmanned quad-copters. The focus is on operations in low-altitude atmospheric conditions, where turbulent gusts are expected to have a significant impact on the performance and stability of small unmanned aerial vehicles. 

Large-eddy simulations (LESs) were performed to accurately represent the atmospheric boundary layer (ABL). The canonical ABL used to generate the data in this study is modeled as a rough flat wall boundary layer with surface heating from solar radiation, forced by a geostrophic wind in the horizontal plane and solved in the rotational frame of reference fixed to the earth's surface. From the LES data, a reduced-order representation of the wind field was also constructed. Additionally, the Dryden turbulence model for wind velocity fluctuations was included as a benchmark wind model. 

Quad-copter aerodynamics was modelled using adaptations of blade element momentum theory. Models for thrust, drag and power of the quad-copter were integrated with the flight dynamics and wind models.  A non-linear flight controller (backstepping controller) was developed  to control all six degrees of freedom of the motion of the quad-copters.  The was studied.

An ascent-straight-descent path and a circular path were designed for the simulations of the closed-loop system. These two trajectories were of interest due to the fact that they both incorporate a representative set of possible trajectories of a quad-copter. 

Representative results for flight parameters, required RPM inputs, resultant trajectory and power of the quad-copter for different aerodynamic and wind models and planned trajectories were obtained and compared against each other. A multiple cruise-speed trajectory phase was defined to determine the optimal cruise speed of the quad-copter.


 
Collectively, this study presented a new suite of tools for realistic, flight simulations, and provides insight into the impact of modeling fidelity on trajectory planning and control. The entire simulation is open-sourced for use by the community~\footnote{\url{ https://github.com/behdad2018/FlightSim_QR_AIAA}}.

\section{Acknowledgments}
This work was supported by NASA under the project 
``Generalized Trajectory Modeling and Prediction for Unmanned Aircraft Systems'' (Technical Monitor: Sarah D'souza).

\section*{Appendix: Validity of momentum theory}

The versatile dynamics of vertical take-off and landing (VTOL) vehicles such as a quad-copter creates different flow regimes surrounding the vehicle. It is instructive to analyze the validity of the assumptions made about the aerodynamics of the quad-copter, which is based on Momentum theory. In rotorcraft aerodynamics, it is well-known that there are multiple structural configurations that the  rotor wake can assume. Specifically, when the ratio of climb inflow ratio to the hover inflow ratio satisfies $-2<\lambda_c/\lambda_h<0$, the momentum theory is not valid. This analysis was carried out only for a rotor. Here, $\lambda_h$ is the hover inflow ratio and is simply given by $\lambda_h=\sqrt{C_\text{T}/2}$. In hover, thrust is equal to the weight ($T=W$). Therefore, $\lambda_h=\sqrt{\frac{W}{4}/(\rho\pi R^2 V_\text{tip}^2})$ which is $0.0784$ for the rotor studied herein. Also, the required RPM in hover is about 10150.

 It is instructive to further analyze the two different paths explored previously and determine how those paths fit into the flight envelope and  momentum theory theory regimes.


\begin{figure}[H]
\centering
\includegraphics[scale=0.45]{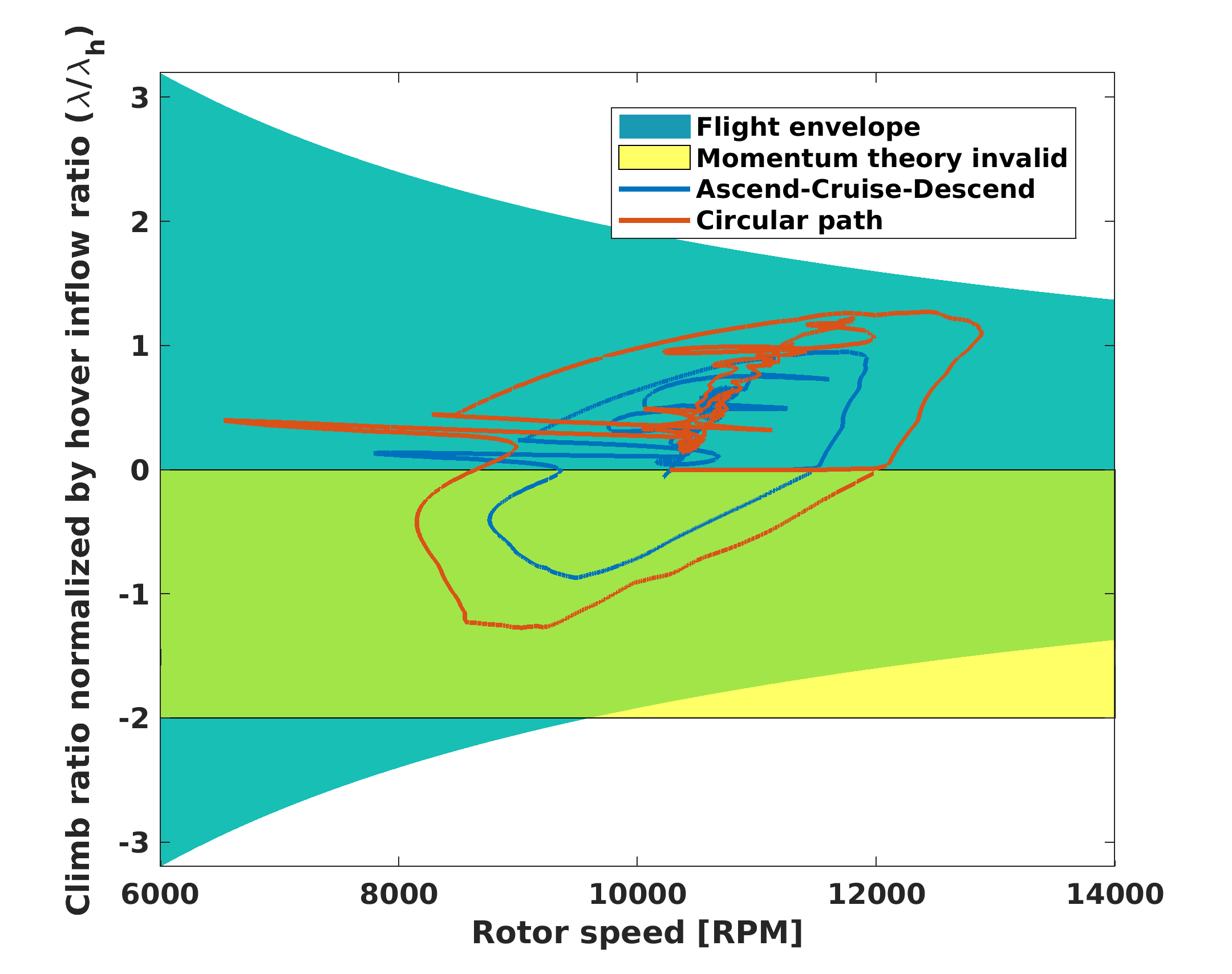}
\caption{Validity of the momentum theory.}
\label{fig:flightenv}
\end{figure}

Fig.~\ref{fig:flightenv} shows that the majority of the both paths are in the valid region, indicating that the use of momentum theory is suitable. It is noted that the portions of the curves that are in the invalid region of momentum theory (shown by shaded yellow) belong to the landing phase of the flight where the rotor is in vortex ring state, and momentum theory is not strictly valid, but might still be used as a rough approximation.

\begin{figure}[H]
\centering
\includegraphics[scale=0.24]{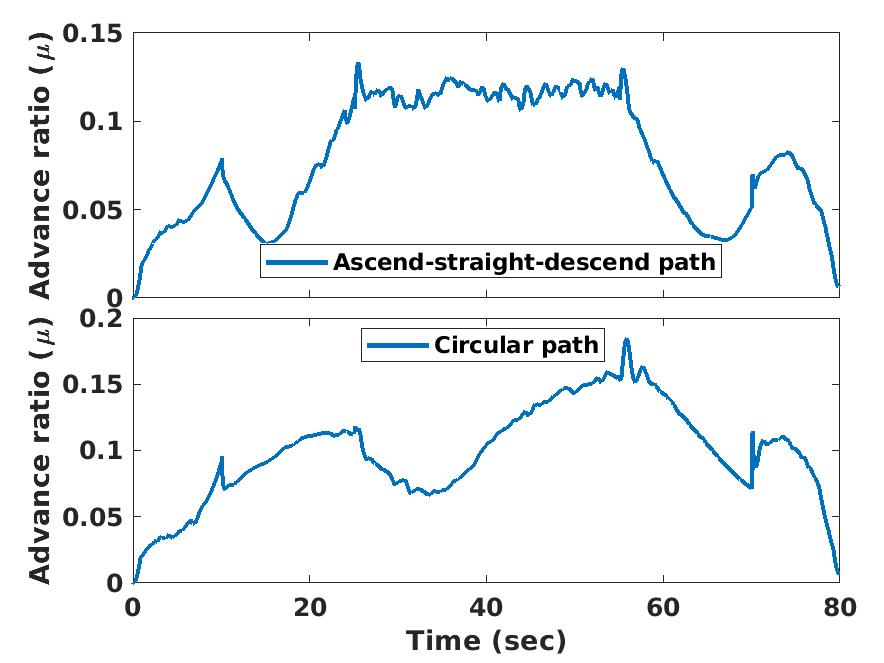}
\caption{Advance ratio for ascent-straight-descent and circular paths.}
\label{fig:mu}
\end{figure}

The advance ratio is shown in Fig. \ref{fig:mu}. Given that the rotor speeds are  high, the relatively low advance ratio values further offer credence to momentum theory.

\section*{References}
\bibliography{ref.bib}
\bibliographystyle{aiaa}

\end{document}